\documentclass[a4paper,11pt]{article}
\pdfoutput=1
\usepackage{jcappub}
\usepackage{graphicx}
\usepackage{hyperref}
\usepackage{xcolor}
\usepackage{amssymb,amsmath,amsthm}

\title{Boosting the dark matter signal with Coulomb resonances}

\author[a,b]{Rakhi Mahbubani}
\author[a]{and Kin Mimouni}

\affiliation[a]{Theoretical Particle Physics Laboratory (LPTP),
  Institute of Physics, EPFL, Lausanne, Switzerland.}
\affiliation[b]{Albert Einstein
  Center for Fundamental Physics, Institute for Theoretical Physics, University of Bern, Sidlerstrasse 5,
  CH-3012 Bern, Switzerland}
\emailAdd{rakhi@cern.ch}
\emailAdd{kin.mimouni@epfl.ch}

\abstract{We show that the presence of nearby Coulombic resonances at finite
  energy could lead to the enhancement of the dark matter annihilation
  cross section at specific non-zero velocities correlated with the
  mass splitting between the dark matter pair and that of the 
  resonance.
If one of these resonant velocities approximately matches the velocity of dark matter
in our local neighbourhood, we would see this
enhancement in existing indirect-detection measurements, such as the
measurements of the continuum photon spectrum made by HESS and Fermi-LAT.
We explore this effect in the context of pure Higgsino and
  Wino dark matter with a variable splitting between charged and
  neutral components,
  controlled by the
  Wilson coefficient of a higher-dimension operator.  
For electroweak WIMPs a relevant and appreciable enhancement from Coulomb
resonances requires tuning the charged-neutral splitting to be of order the Coulomb binding
energies.   
This leads to strong exclusions of Higgsino dark matter
with charged-neutral splittings in the narrow ranges (2, 2.5) and
(8.5, 10.5) MeV.  
In contrast, by decreasing the charged-neutral splitting for the
thermal Wino, we can move the Yukawa resonance away from the thermal
relic mass, decreasing the indirect-detection signal to a level that
is compatible with HESS measurements in the window (25, 35) MeV.}

\begin{document}
\maketitle

\section*{Introduction}

There is compelling astrophysical and cosmological evidence for the
existence of Dark Matter (DM) that calls for physics beyond the
Standard Model (SM).  The simplest solution is the Weakly-Interacting
Massive Particle (WIMP) paradigm.  This consists of a new
weakly-interacting state whose interaction with the SM freezes out of thermal equilibrium,
naturally yielding the measured dark matter density for mass around a
TeV, the scale by which we also already expected to have seen the New
Physics that solved the hierarchy problem.  The slow erosion of the
latter hope by null results at the LHC has devalued the currency of
the WIMP paradigm, opening up the parameter space for viable dark
matter to both extremely light and extremely heavy scales,
and resulting in a frenzy of theoretical and experimental activity
at the boundary of condensed matter and astroparticle physics (see
e.g. \cite{Battaglieri:2017aum} and references therein).
Nevertheless the original WIMP solution lives on - its small cross
section making it hard to probe at hadron colliders, particularly
for a fairly pure Higgsino with loop-induced splittings,
and greater ingenuity \cite{Baryakhtar:2017dbj,Krall:2017xij} (or perhaps future colliders \cite{Low:2014cba,Mahbubani:2017gjh}) will be
required to discover or exclude it.

A large separation of scales between the mass of the
dark matter particle and that of its force carriers, could lead to a non-perturbative enhancement of its annihilation cross
section in the nonrelativistic limit due to multi- gauge boson exchange.  This Sommerfeld
enhancement, first documented in the context of multi-photon
exchange by electrons \cite{Sommerfeld:1931}, was later demonstrated
\cite{Hisano:2004ds,Hisano:2006nn} to affect the mass and indirect detection cross section of a
pure Wino thermal relic.  The significant enhancement of the latter
can be seen as due to the accidental presence of a `zero-energy' resonance close
to the DM-DM ground state at the relic mass, loosely bound by the
attractive Yukawa potential between the neutral particles \cite{Hisano:2004ds}.  This is
not the case for the Higgsino relic, the Yukawa potential having no bound state for DM mass around the relic mass.

In addition to a potential global enhancement at low velocity due to
zero-energy bound states of dark matter, there can
also be a significant enhancement of the dark matter annihilation
cross section at {\it specific} velocities
due to quasi-bound states of a nearby, more massive two-particle
state.  For simplicity, in this work we will illustrate this behaviour using pure Higgsino and pure
Wino dark matter as simple test cases.  In these examples electroweak loop corrections
lift the mass of the charged component with respect to the neutral one
by $\delta m\sim \alpha m_Z$, and gauge boson exchange allows mixing
between the neutral-neutral (DM$^0$DM$^0$) and heavier charged-charged
(DM$^+$DM$^-$) two-particle states.
The dark matter spectrum will also contain various two-particle states that
are loosely bound by the long-range interactions.  The lowest-lying
state will be the DM$^0$DM$^0$ quasi-bound state
responsible for the large boost in the Wino annihilation cross-section.  More central to our argument, however, is the
tower of Coulomb DM$^+$DM$^-$ resonances that lie between
the free DM$^0$DM$^0$ and DM$^+$DM$^-$ two-particle states.  When
the incoming dark matter particles have just enough energy to create
these states, their annihilation cross section is resonantly enhanced;this
enhancement only occurs for certain specific values of the incoming
velocity that match the mass difference between the ground state and
the resonance.

For pure Wino/Higgsino states with nominal splitting, the Coulomb
binding energies $E_{B,\gamma}=\alpha^2M_\chi/(4n^2)$ are negligible
compared with the splittings, so the resonant
velocities ($\beta\sim 10^{-2}$) are irrelevant for any measured
physical processes.  Nevertheless in any scenario where the Coulomb
binding energy is of the order of the splitting, the lowest-lying resonances could be brought
close enough to the ground state to give a large boost to the
indirect-detection signal for dark matter at velocities relevant to
existing measurements in our galactic centre (GC) and/or nearby dwarf
spheroidal galaxies (DSG).  Hence varying the charged-neutral
splitting, via a higher-dimension operator originating from
integrating out heavy new physics for example, could give rise to
significant changes in the cosmology and phenomenology of the dark
matter candidate by tuning Yukawa and Coulomb resonances in and out of relevant regions of phase space.

We begin this work with a review of Sommerfeld enhancement in Section
\ref{sec:Sommerfeld}, building some intuition for this phenomenon
using simple scenarios where it can be computed analytically.  We
continue in
Section \ref{sec:Models} with an exposition of our simplified model
framework, consisting of the SM with the
addition of a weak-doublet (triplet) fermion with hypercharge
$1/2$ $(0)$ and Dirac (Majorana) mass, corresponding to the pure Higgsino (Wino) limit of the
Minimal Supersymmetric Standard Model (MSSM). We assume
that the charged-neutral splitting is a free parameter, varied using the
Wilson coefficient of a higher-dimension operator.  We consider the
effect of varying the splitting on the indirect-detection cross
section in Section \ref{sec:IndirectDetection}, focussing on the
continuum photon spectrum which does not suffer from the large
propagation uncertainties of hadronic final states.  We compare the results
with existing measurements from HESS and Fermi-LAT.  Finally we set
our simplified model within the context of the MSSM, and the MSSM with
Dirac gauginos in Section \ref{sec:FullModels}.  We present our
conslusions and outlook in Section \ref{sec:Conclusion}.


\section{Sommerfeld enhancement}
\label{sec:Sommerfeld}
This is the enhancement of the short-distance cross section for a
process due to the distortion of the wave function for the incoming
state by a long-range potential.  This effect increases with
decreasing velocity, and was first noticed by
Sommerfeld \cite{Sommerfeld:1931} in the context of electromagnetism.  It has a classical gravitational analogue in the low-velocity
enhancement of the cross section for a point particle hitting a
massive object of radius R \cite{ArkaniHamed:2008qn}:
\begin{equation}
\sigma=\pi R^2\left(1 + \frac{\beta_{\textrm esc}^2}{\beta^2} \right)
\end{equation}
where $\beta$ is the velocity of the point particle and $\beta_{\textrm esc}=2
G_N M/R$ is the escape velocity from the surface of the extended
object.

In quantum field theory it can be seen as the enhancement due to
ladder diagrams in which a light force-carrier is exchanged.  For
non-relativistic muon annihilation for example, the amplitude for the n$^\textrm{th}$-order ladder
diagram (due to n photon exchanges between the muon pair) is
proportional to $(\alpha/\beta)^n$ in the non-relativistic limit, which
means the perturbative expansion in $\alpha$ breaks down for small
enough velocity, and the ladder diagrams must be systematically
resummed.  As shown in \cite{Hisano:2006nn} the
Sommerfeld factor can be determined by factorizing the short-distance
and long-distance behaviour.  The distorting effect of the long-range
potential $V(r)$ on the two-particle wavefunction is computed by solving the
Schr\"{o}dinger equation in the presence of only this potential, and the total enhancement then computed by
perturbing around the resulting inhomogeneous solution at leading order in the
absorptive part, which encodes the short-distance behaviour.  The same
procedure was proved in \cite{Iengo:2009ni} to be equivalent to
explicit resummation of the ladder diagrams.

Two particles subject to a long-range potential $V(|\vec{r}|)$,
satisfy the following Schr\"{o}dinger equation in the centre-of-mass
frame, where $\psi(\vec{r})$ is the two-particle wavefunction:
\begin{equation}
\frac{-1}{2\mu}\nabla^2\psi(\vec{r})+ V(|\vec{r}|)\psi(\vec{r})=\frac{1}{2}\mu \,\dot{\vec{r}}\,^2 \psi(\vec{r})
\end{equation}
with $\mu$ is the reduced mass of the system.

Separating variables to isolate the radial and angular parts as usual,
we obtain, for two particles of equal mass $m_1=m_2=M$, with 
velocity $\beta$ in the centre-of-mass frame\footnote{We use the velocity $\beta$ such that in the non-relativistic limit we have the Mandelstam variable $s=4M^2 + 4 M^2 \beta^2 + O(\beta^4)$}, the following radial equation:
\begin{equation}
-\frac{1}{M}\frac{1}{r^2}\frac{d}{dr}\left(r^2\frac{dR}{dr}\right) + \left(\frac{l(l+1)}{Mr^2}+V(r)-M\beta^2\right)R=0
\end{equation}
To isolate the dominant ($s$-wave) contribution, we take $l=0$; substituting $R(r)=\chi(r)/r$ we obtain
\begin{equation}
\chi''(r)+\left(M^2 \beta^2 -M\, V(r)\right)\chi(r)=0\;.
\label{equ:1DSommerfeld}
\end{equation}
This equation cannot be solved analytically for general potentials
$V(r)$, but we can solve numerically for
the irregular solution, which satisfies the boundary conditions:
\begin{equation}
\chi(0)=1,\,\chi'(\infty)=-i M\beta \chi(\infty)
\end{equation}
and then compute the Sommerfeld-enhanced cross section as
\begin{equation}
\sigma\equiv S \,\sigma_0=\frac{|\chi(\infty)|^2}{|
\chi(0)|^2} \sigma_0
\end{equation}
where $\sigma_0$ is the perturbative cross-section and $S$ denotes the
Sommerfeld enhancement factor.\footnote{This procedure was shown in \cite{ArkaniHamed:2008qn} to be equivalent to finding
the regular solution and computing the enhancement using
\begin{equation}
S=\left|\frac{\frac{d\chi}{dr}(0)}{k}\right|^2
\end{equation}}

This method can be generalized to the case where there is an
interaction that mixes distinct two-particle states.  For $N$
two-particle states $\chi_i(r)$,
with $i=1,\cdots,N$, the Sommerfeld enhancement can be computed by
numerically solving the following set of $N$ coupled radial Schr\"{o}dinger
equations:
\begin{equation}
\chi_i''(r)+M^2\beta^2\chi_i(r)-M\sum_{j}\ V^\textrm{tot}_{ik}(r)\chi_k(r)=0
\end{equation}
$N$ times, with a different boundary condition at $r=0$ each time,
representing each distinct two-particle state participating in
the short-distance interaction: $\chi_i(0)|_j=\delta_{ij}$ for the
$j^\textrm{th}$ solution.  The boundary condition at infinity
corresponds to a pure outgoing (decaying) wave for
$(M\beta^2-V(r\to\infty))$ positive (negative).  Here all energies/potentials are defined with respect to the lightest
two-particle state, consisting of two identical particles of mass
$M$. Any mass differences with respect to the lightest state show up
as additional radius-independent contributions to the total
potential.\footnote{Note for states $\chi_i$ with mass $M+\delta m_i$,
  $\beta$ is \emph{not} the physical initial velocity of the particle,
  but is defined such that the total energy of the heavy-particle pair with respect
  to the dark-matter pair is $E=M\beta^2$.  The Sommerfeld
  enhancement calculation takes into account that the state $\chi_i$ cannot
  exist as an asymptotic state for $M\beta^2 < \delta m_i$.
}

The cross-section in the $i^\textrm{th}$ channel is then given by:
\begin{equation}\label{eq:SFcross}
\sigma_i=c_i(A\,.\,\Gamma\,.\,A^\dag)_{ii}
\end{equation}
where $\Gamma_{jk}$ is the absorptive part of the two-to-two cross
section $\chi_j\to\chi_k$, with $j,k$ running over all possible two-particle
states in each channel.
$c_i$ is a numerical factor that accounts for the different
normalization of the two-body wavefunctions for identical and
non-identical particles: $c_i=2$ ($c_i=1$) when the two particles are
identical (distinct).  The matrix $A$ is computed as:
\begin{equation}\label{SFdef}
A_{ij}=\lim_{r\to\infty}\frac{\chi_i(r)|_j}{e^{i
    \sqrt{M(M\beta^2-V_{ii}(\infty))} \,r}}\;.
\end{equation}
We can also define a Sommerfeld factor in each channel $i$, in analogy with the unmixed
case:
\begin{equation}
\label{eq:SommerfeldComponents}
S_i=c_i\frac{(A\,.\,\Gamma\,.\,A^\dag)_{ii}}{\Gamma_{ii}}.
\end{equation}
Note that unlike in the unmixed case the mixed Sommerfeld factor is not only
dependent on the potential, but also on the $\Gamma$ matrix for the
hard process of interest.

In practice, for mixed channels we must solve for the Sommerfeld factor
numerically.  The numerical stability of the solution is an issue in
the multi-state scenario, since the presence of an exponentially
decaying part in the charged-charged wavefunction (due to the
charged-neutral mass difference) makes the solution particularly
unstable to rounding errors.  Special care must be taken in the
numerical recipe in order to favour convergence. We use the variable
phase method, as detailed in \cite{Beneke:2014gja}.

\subsection{Some analytical considerations}
The electroweak potentials that we will consider have three
individual constituents that combine in a complex way: the Coulomb
interaction, the Yukawa interaction due to weak gauge boson
exchange, and a mass splitting between different sets of asymptotic
states.  The effect on the Sommerfeld factor of each of these
components individually can be understood analytically.  Following
\cite{Slatyer:2009vg}, we will review these arguments below in order
to build some intuition before we move on to tackle the full
electroweak case in Section \ref{sec:EW} below.

\subsubsection*{Pure Coulomb potential} 
We first consider the scattering of a pair of particles of mass $M$
and velocity $\beta$ in their centre-of-mass frame
interacting via a Coulomb potential $V(r) = \pm\alpha /
r$, where the Sommerfeld factor can be computed analytically \cite{ArkaniHamed:2008qn,Mitridate:2017izz}:
\begin{equation} \label{SEEM}
S = \left|\frac{\mp\frac{\pi \alpha}{\beta}}{1-e^{\pm\frac{\pi\alpha}{\beta}}} \right|.
\end{equation}
The enhancement
is determined by the relative significance of the Coulomb binding energy
$\alpha^2 M$ of the incoming particles and their kinetic energy of,
$M\beta^2$, and is large when the binding energy dominates.  Hence the
Sommerfeld factor is independent of the particle $M$, and grows with
$\alpha/\beta$ for an attractive potential, approaching unity for large
velocities as expected (in practice the arbitrary growth at small
$\beta$ will be cut off
by finite temperature effects, when the photon acquires a thermal
mass). For a repulsive potential it gives rise to an exponential
suppression, due to the presence of a Coulomb barrier:
\begin{equation}
\lim_{\beta\to 0}S\sim\left\{
\begin{array}{l l}
\frac{\pi \alpha}{\beta} & \hspace{1in}\textrm{attractive}\,V \\
\frac{\pi\alpha}{\beta}e^{-\frac{\pi\alpha}{\beta}} & \hspace{1in}\textrm{repulsive}\,V
\end{array}\right.
\end{equation}
%

\subsubsection*{Pure Yukawa potential} 
The enhancement due to a Yukawa potential
$V(r)=\pm\frac{\alpha}{r}e^{-M_V r}$, arising from the exchange of a
massive gauge boson of mass $M_V$, can only be treated analytically
by approximating the Yukawa potential as a Hulth\'{e}n potential:
\begin{equation}
 V(r)=\pm\alpha \frac{k M_V e^{-k M_V r}}{1-e^{-k M_V r}}
\end{equation}
where $k$ is a fudge factor chosen phenomenologically to match the
short- and long-range behaviour of the Yukawa potential ($k=\frac{\pi^2}{6}$ for $s$-wave
enhancement \cite{Cassel:2009wt}). The Sommerfeld factor can be
expressed in closed form \cite{Cassel:2009wt,Slatyer:2009vg,Cirelli:2016rnw}:
\begin{equation}\label{SEHulthen}
S = \mp\frac{\pi \alpha}{\beta} \frac{\sinh \left(\frac{2 \pi M}{k M_V}  \beta \right)}{\cosh \left(\frac{2 \pi M}{k M_V}  \beta \right) - \cos \left(\frac{2\pi M}{k M_V}  \beta \sqrt{\mp\frac{k M_V}{M}\frac{\alpha}{\beta^2}-1} \right)}
\end{equation}
The Yukawa form of the potential brings into play another relevant
quantity: the size of the Bohr radius, $(\alpha M)^{-1}$ relative to the
range of the potential $M_V^{-1}$; the Sommerfeld enhancement grows with
$\alpha M/M_V$:
\begin{equation} \label{slowYukawa}
\lim_{\beta\to 0}S \sim \left\{
\begin{array}{l l}
2k\left(\frac{\alpha}{\beta}\right)^2\frac{ M_V}{\alpha M } &
                                           \hspace{0.2in}\textrm{attractive}\,V,\;M=n^2
                                           \frac{k M_V}{\alpha}
                                           \quad\textrm{for integer }
                                           n\\
\\
\frac{2\pi^2}{k} \frac{\alpha M}{M_V} \left(1- \cos \left(2\pi \sqrt{\frac{\alpha M}{k M_V}}\right)\right)^{-1}&
                      \hspace{0.2in}\textrm{attractive}\,V,\;\textrm{non-resonant
  M}\\
\\
\frac{2\pi^2}{k} \frac{\alpha M}{M_V}\left(\cosh \left(2\pi
  \sqrt{\frac{\alpha M}{k M_V}} \right)-1\right)^{-1} & \hspace{0.2in}\textrm{repulsive}\,V
\end{array}\right.
\end{equation}
Unlike that for the Coulomb potential, the Sommerfeld factor for an
attractive Yukawa potential tends to a constant as $\beta$ approaches
zero, except for specific values of $M$ where the denominator of
\eqref{slowYukawa} vanishes, and the Sommerfeld factor undergoes even
faster growth, as $1/\beta^2$.  These values of $M$ correspond to
instances where the Hulth\'{e}n potential admits zero-energy bound
states.  For a repulsive Yukawa potential, the Sommerfeld factor again
tends to a constant, that smaller than one, and there is no resonant structure.

\begin{figure}[h]
\begin{center}
\includegraphics[width=0.48\textwidth]{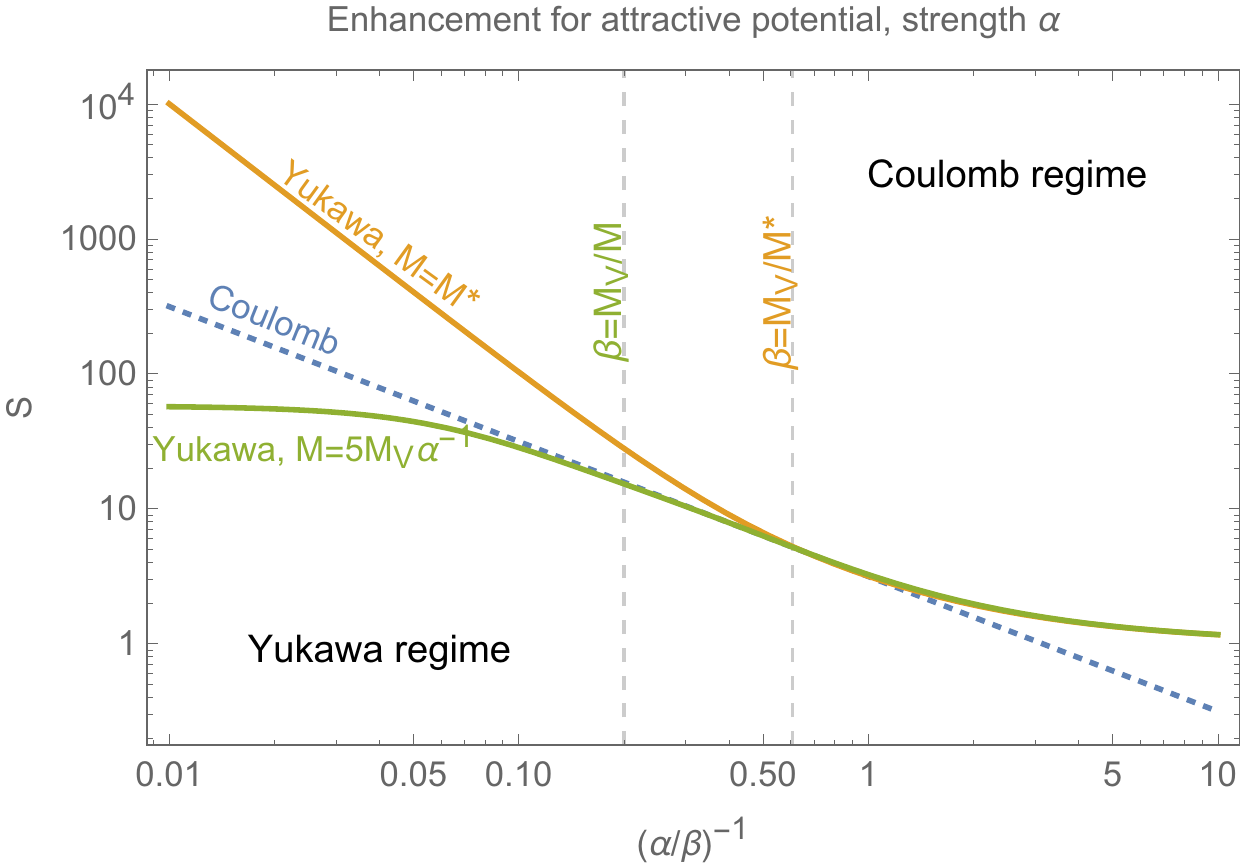}
\end{center}
\caption{Sommerfeld enhancement for an attractive Yukawa potential,
  showing both off-resonance saturation for $M_V(\alpha M)^{-1}=1/5$
  (solid green curve), and
  on-resonance growth for $M=M^*=k n^2
  M_V/\alpha$ (solid yellow curve).  The enhancement for a pure
  Coulomb potential is shown for comparison (dashed blue line).  For further explanation of
  symbols, see text. }
\label{fig:AnalyticalSommerfeldYuk}
\end{figure}

For larger values of $\beta$, the hyperbolic functions simplify and
$S\sim \pi \alpha/\beta$ as in the electromagnetic case. The
transition between the two regimes happens around $\beta \sim M_V/M$
\cite{ArkaniHamed:2008qn}, where the de Broglie wavelength of the
particle becomes of order the range of the Yukawa interaction, and the
particle begins to probe the short-range nature of the potential.  See
Figure \ref{fig:AnalyticalSommerfeldYuk} for a graphical display of
the behaviour of Sommerfeld factor for an attractive Yukawa potential as a function
of $\beta/\alpha$, for $M_V(\alpha M)^{-1}=0.2$ (solid green curve),
as compared with the growth for an on-resonance mass, $M=M^*=k n^2
  M_V/\alpha$ (solid yellow curve).

\subsection*{Mass splitting}

The introduction of a mass splitting $\delta m$ between the incoming state and a
nearby state with which it mixes has two important consequences.
Above some threshold velocity 
\begin{equation}\label{EMThresh}
\beta_{\textrm{th}}=\sqrt{2\frac{\delta m}{M}}\; ,
\end{equation}
a pair of incoming states can scatter inelastically to a pair
of heavy partners on-shell, giving the light states access to a new,
perhaps stronger, annihilation channel.  More crucial to our
narrative, however, are the large enhancements that can occur at
particular velocities below
the heavy-particle threshold, due to threshold production of loosely-bound resonances of the
heavy partner pair, lying between the light and heavy states in the
mass spectrum.  For a resonance with binding energy $E_B$, the
corresponding resonant velocity can be written as:
\begin{equation}
\beta^*=\sqrt{\frac{2\delta m-E_B}{M}}
\end{equation}
The physically-relevant parameter encoding the splitting is then
$\delta m/E_B$.  The enhancement due to these below-threshold resonances was observed in
\cite{Slatyer:2009vg,Asadi:2016ybp,Braaten:2017dwq}, but their
possible effects on the phenomenology of dark matter were not
explored.

\subsection*{Mixed electroweak potential}
We will be concerned with a two-component non-relativistic potential,
due to mixing between the charged (DM$^+$DM$^-$) and neutral (DM$^0$DM$^0$) components of the dark matter
multiplet, that takes the following schematic form:
\begin{equation}
 V=
 \begin{pmatrix}
 \delta m + V_C + V_Y & V_Y \\ V_Y & V_Y 
 \end{pmatrix}
\end{equation}
where $V_C,\,V_Y$ denote the Coulomb and Yukawa potentials, with
coupling strength scaling like $\alpha$ and $\alpha_L$, respectively.
The combined effect on the Sommerfeld factor will result in different
behaviour in different velocity regimes, as follows:
\begin{itemize}
 \item High $\beta$, $\beta \gg \frac{m_W}{M},\, \beta_{th} $: in this
   regime we can neglect the mass splitting and trust the analytical
   expressions for the Sommerfeld factor. The Yukawa contribution is of order $S_{L}\sim \pi \alpha_L/\beta$ and the electromagnetic contribution is of order $S_\textrm{em}\sim \pi \alpha/\beta$ and since $\alpha_L > \alpha$ we expect the Yukawa potential to dominate.
 \item Intermediate $\beta$, $\frac{m_W}{M} > \beta >
   \beta_\textrm{th} $: in this regime we can also neglect the mass splitting. Off resonance, the Yukawa contribution goes to a constant of order $S_{L}\sim \alpha_L M/m_W$  while the electromagnetic contribution continues to grow like $S_\textrm{em}\sim \pi \alpha/\beta$, so we expect the electromagnetic attraction to dominate. On resonance however, the Yukawa contribution behaves like $1/\beta^2$ and is the dominant one.
 \item Low $\beta$, $\beta < \beta_\textrm{th}$: the $\delta m$ term
   in the potential dominates. In charged channels, the Sommerfeld
   factor is zero by construction because an initial on-shell charged pair cannot have an energy below $2(M+\delta m)$. Just below threshold the neutral channels
   exhibit resonant enhancement due to Coulomb resonances of
   DM$^+$DM$^-$. As $\beta$ decreases still further, the weak Yukawa potential dominates.
 \end{itemize}
We will see these different types of behaviour in the context of pure
Higgsino and Wino simplified models in Section \ref{sec:Models} below.


\section{Simplified electroweak models}
\label{sec:Models}

We will explore the Sommerfeld enhancement due to the partner resonances in the
framework of two simplified models, corresponding to the
pure Higgsino and pure Wino limits of the MSSM, with the addition of the leading higher-dimension
operators that affect the tree-level mass spectrum in each case.  Our
arguments will be placed within the context of the MSSM and the MSSM
with Dirac gauginos in Section \ref{sec:Models}.

Our first example consists of the Standard Model with the addition of a
vector-like $SU(2)_L$ doublet $\chi$ of fermions. In order to have an
electrically-neutral thermal relic candidate, $\chi$ must have
hypercharge $Y=\frac{1}{2}$ and thus be a Dirac doublet. The
Lagrangian for $\chi=(\chi^+, \chi^0)$ is:
\begin{equation}\label{SimplifiedHino}
\mathcal{L}_\textrm{doublet}=\mathcal{L}_\textrm{SM} + i\bar \chi D \!\!\!\!/ \chi - M \bar \chi \chi+ c_1 H^\dagger H \, \bar \chi \chi + c_2 H^\dagger \frac{\sigma^a}{2} H \, \bar \chi \frac{\sigma^a}{2} \chi +\cdots
\end{equation}
where we have included the leading corrections to the dark matter mass
due to new physics, in the form of dimension-5 operators
with arbitrary Wilson coefficients $c_1$ and $c_2$.  If the new
physics has coupling $g_*$ and mass $M_*$, we would expect these
coefficients to be $\mathcal{O}(g_*^2/M_*^2)$, up to an order-one
constant.  

We enforce stability of the DM candidate $\chi^0$ by imposing a global symmetry that preserves DM number (in the MSSM it is $R$-parity), forbidding terms such as $\bar L \chi$ in the Lagrangian (where $L$ is the lepton doublet). 

Setting the Higgs to its vacuum-expectation value (VEV), we immediately see that the first term
is a global shift of the charged and neutral DM mass, which we can
redefine away, whereas the second term gives rise to the following mass splitting
between the states:
\begin{equation}\label{MDM2split}
\delta m = M_{\chi^+} - M_{\chi^0} = \frac{c_2 v^2}{2} = \frac{c_2 m_W^2}{2g_L^2}
\end{equation}
This splitting can be positive or negative, depending on the sign of
$c_2$, and could be as large as 2 GeV for a new physics state of mass $M_* \sim 4$~TeV with weak coupling $g_* \sim 1$. 
There is also a contribution to the splitting coming from electroweak
loops.  This is due to different wave-function renormalization for the neutral and
charged state from a photon and $Z$-boson loop, and takes the form
\cite{Thomas:1998wy}:
\begin{equation}
\delta m = M_{\chi^+} - M_{\chi^0} = \frac{\alpha M}{2\pi} \int_0^1 dx\, (1+x) \log \left(1+\frac{1-x}{x^2} \frac{m_Z^2}{M^2} \right)
\end{equation}
where $M$ is the DM mass. In the limit we are interested in, namely $M
\gg m_Z$ this integral reduces to $\delta m = \frac{\alpha m_Z }{2\pi}
= 344$ MeV.

Because it has a non-zero hypercharge, the doublet has a direct
coupling to the $Z$ boson and a scattering cross-section on nuclei
around $10^{-39}$~$\mathrm{cm}^2$, several orders of magnitude larger
than the limits imposed by direct detection experiments
\cite{Aprile:2018dbl}.  A way around this is to assume that the
neutral state $\chi^0$ has a mass mixing with a neutral Majorana state
that splits the Dirac fermion into two Majorana fermions $\chi_1$ and
$\chi_2$. The lightest state $\chi_1$ is the DM candidate and has no
diagonal $Z \chi_1 \chi_1$ coupling. This scenario occurs naturally in
the MSSM where the neutral Higgsino components mix with the Bino and
neutral Wino. Then the direct detection process
goes through the off-diagonal $Z$-coupling and is the inelastic
reaction: $\chi_1 N \rightarrow \chi_2 N$. If $M_{\chi_2} - M_{\chi_1}
\ge 350$ keV or so (depending on the nucleus), this process is
kinematically forbidden and cannot be seen in direct detection
experiments \cite{Bramante:2016rdh}. In the remainder of this work we will be
assuming that such a neutral splitting is present, and is large enough
to avoid direct detection constraints, but has a negligible effect on
our Sommerfeld calculations. 

Our second simplified model consists of the SM
lagrangian plus a $SU(2)_L$ triplet $\chi^a$ of Majorana fermions with zero
hypercharge, as follows:
\begin{equation}
\mathcal{L}_\textrm{triplet}=\mathcal{L}_\textrm{SM} + i (\chi^a)^\dagger \bar
\sigma_\mu \partial^\mu \chi^a - \frac{M}{2} \chi^a \chi^a
+\mathcal{L}_5+c_3 \left(H^\dagger \frac{\sigma^a}{2} H
  \chi^a\right)^2 + \textrm{h.c.} + \cdots
\end{equation}
As in the doublet case we assume a global symmetry which ensures the
stability of the DM state, $\chi^0$.  Here we have omitted the dimension-5 contributions to the triplet mass
which only give rise to a global shift of the masses, while keeping the first non-trivial
contribution to the mass splitting, which arises only at dimension 7.
We would naively expect $c_3$ to scale as $g_*^4/M_*^3$, resulting in
a mass splitting:
\begin{equation}
\delta m = M_{\chi^\pm } - M_{\chi^0} = 2 c_3 \frac{m_W^4}{g_L^4}
\end{equation}
Taking weak coupling and new physics at the scale $M_*\sim 4$ TeV as
before, the tree-level contribution to the splitting is tiny, $\delta
m\sim 7$ MeV, in comparison with the contribution coming from
electroweak loops.  In the limit $M \gg m_Z$ the latter yields $\delta m=\alpha_L m_W \sin^2 \frac{\theta_W}{2} = 165$
MeV \cite{Cirelli:2005uq}.  A large change in the splitting away from
the electroweak loop value would therefore require strong coupling at
a relatively low scale.

For the purposes of this work we will treat $\delta m$ as a free
parameter, allowing it to vary freely in the range $[0, \,
1]$ GeV.  For reasons that will become clear later we will
be most interested in the regime where the total splitting is smaller
than the one-loop correction.  This requires a precise cancellation
between the tree-level new physics contribution to the splitting and
the one-loop electroweak correction, which in principle are
parametrically unrelated, constituting a tuning.  We will always require $\delta m$ be positive, in order to have a good dark
matter candidate.  Note that in an abuse of nomenclature we will refer
to the above two simplified models as the `pure Higgsino' and `pure
Wino' scenarios; we thank the reader to keep in mind that we are
allowing for an arbitrary charged-neutral splitting $\delta m$.

For completeness we collect the relevant annihilation matrices and
non-relativistic potentials
for the weak doublet and triplet simplified models in Appendix
\ref{app:Gamma}.  Many of these were taken
from \cite{Hisano:2004ds,Cirelli:2005uq,Cirelli:2007xd}.

\subsection{Electroweak Sommerfeld factors} 
\label{sec:EW}
We will be particularly interested in the Sommerfeld enhancement in
the spin-0, charge-zero channel, which is the only relevant channel
for computing the dark matter indirect-detection cross section.  The
nonrelativistic potential in this channel contains a mixing term between the charged-charged ($\chi^+\chi^-$) and neutral-neutral
($\chi^0\chi^0$) states:
\begin{eqnarray}
 V_\textrm{doublet} &=& \left(\begin{array}{cc}
 2\delta m - \frac{\alpha}{r} - \frac{\alpha_L}{r} \frac{(2c_W^2-1)^2}{4c_W^2} e^{-M_Z r}  &
                                                                    -\frac{\alpha_L}{2r}e^{-M_W
                                                                                             r}\\
\nonumber\\
-\frac{\alpha_L}{2r}e^{-M_W r} & -\frac{\alpha_L}{4c_W^2r} e^{-M_Z r}
\end{array}\right)\nonumber\\
\\
\nonumber\\
V_\textrm{triplet} &=& \left(\begin{array}{cc}
2\delta m - \frac{\alpha}{r} - \frac{\alpha_L c_W^2}{r} e^{-M_Z r}& -\sqrt{2}  \frac{\alpha_L}{r} e^{-M_W
                               r}\nonumber\\ 
\\
-\sqrt{2} \frac{\alpha_L}{r} e^{-M_W r} & 0
\end{array}\right)
\end{eqnarray}
At small velocities, $\beta \ll M/m_{W,Z}$, we would expect the
Sommerfeld factor in the neutral channel to show pure Yukawa-type behaviour, growing like
$1/\beta^2$ at masses where there is a zero-energy resonance and saturating to a
constant away from the resonance as in Equation \eqref{slowYukawa}, with the size of the enhancement
set by the distance from the resonance.  

\begin{figure}[htb]
\begin{center}
\includegraphics[width=0.48\textwidth]{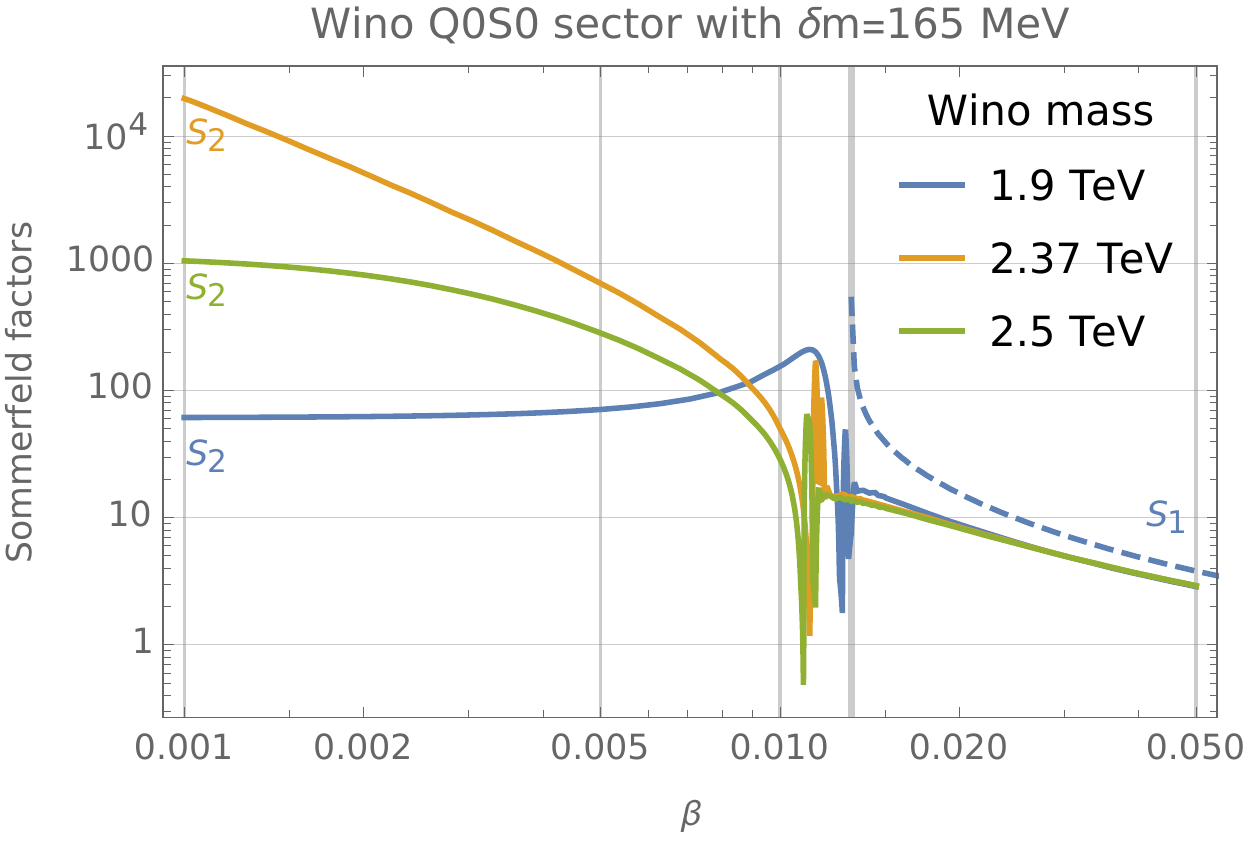} 
~
\includegraphics[width=0.48\textwidth]{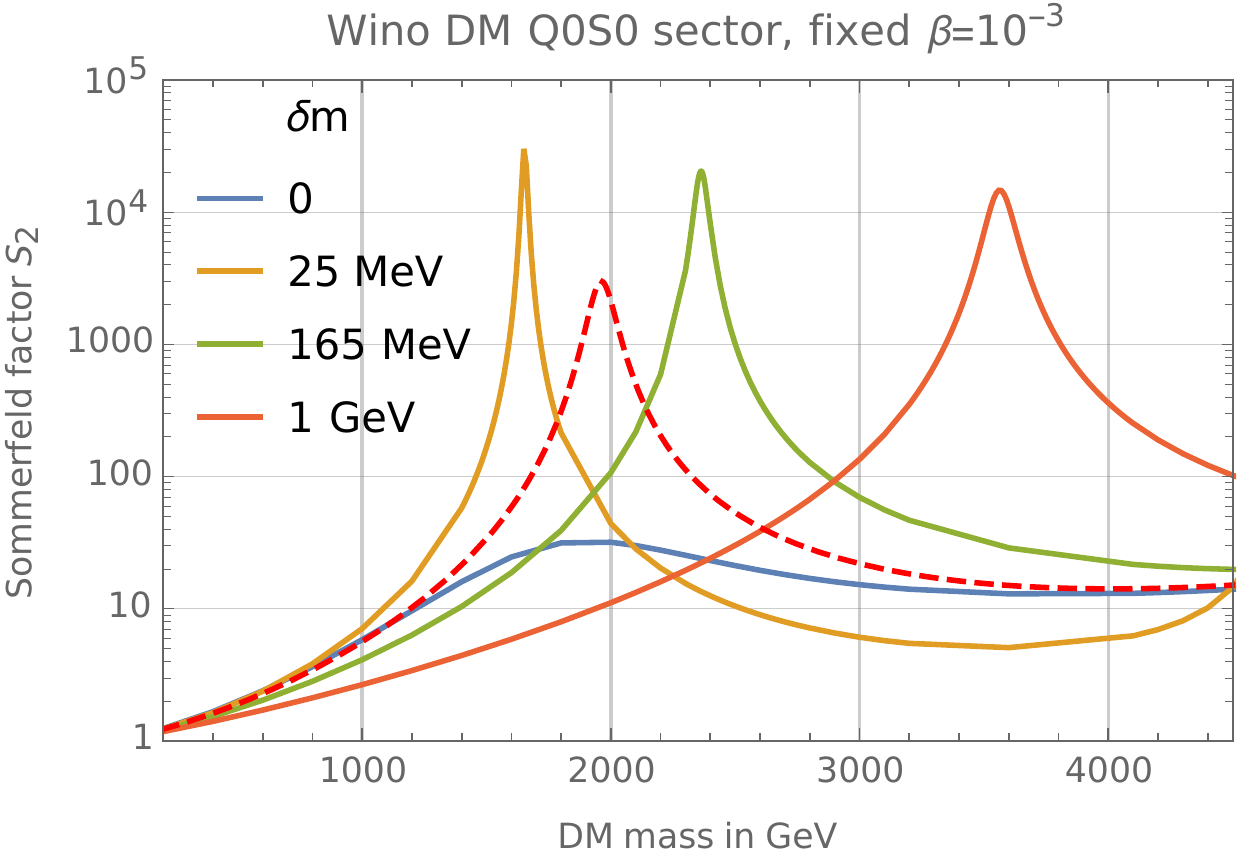} 
\end{center}
\caption{Sommerfeld factors for a pure Wino state. Left panel:
  Sommerfeld factors in the neutral (solid) and charged (dashed) channels
  as a function of DM velocity for various masses around to $M^*\sim
  2.37$ TeV, the position of
  the first zero-energy bound state. The charged-neutral splitting
  $\delta m$ is fixed at the nominal electroweak-loop value, setting
  the position of the Coulomb resonances around $\beta\sim 0.01$.  Right
  panel: Sommerfeld factor in the neutral channel computed at fixed
  velocity ($\beta= 10^{-3}$) as a function of Wino mass
  for varying splitting charged-neutral $\delta m$. We see that the
  resonant mass $M^*$ at which there is a zero-energy bound state
  varies with $\delta m$. We also show the analytical result for
  $g_Y\to 0$ and zero splitting for comparison (dashed line).}
\label{WRes}
\end{figure}

We will begin by examining this effect in the context of the more
familiar Wino case.  In the left panel of Figure \ref{WRes}, we plot numerical results for the Wino Sommerfeld factors
in the charged-charged and neutral-neutral channels ($S_1$ and $S_2$,
respectively, as defined in
Eq. (\ref{SFdef}) and computed for the total annihilation matrix)
as a function of the relative velocity of the incoming neutral states,
$\beta$, for Wino mass varying around the resonant value and nominal
splitting.  Since the velocity is defined with respect to the neutral
states, the charged-charged Sommerfeld factor is zero
below the pair-production threshold for the charged state, $\beta_\textrm{th}$, as
explained in \ref{sec:Sommerfeld} and Equation \eqref{EMThresh}.  We expect the enhancement in this
channel to grow with decreasing $\chi^+$ velocity due to the Coulomb potential; this
growth will be cut off by the non-Coulomb components of the interaction.  

Also visible in this figure is the
enhancement at specific velocities just below to the charged particle
threshold, due to production at threshold of Coulomb `bound states' of
a charged DM pair.  Approximating the binding energy $E_n$ of the
$n$th bound state as being purely due to the Coulomb interaction to
leading order, we expect to see corresponding peaks in the Sommerfeld
factor in the neutral channel at velocities:
\begin{equation} \label{EMBS}
\beta^*_{n}=\sqrt{\frac{2\delta m}{M}-\frac{\alpha^2}{4n^2}}
\end{equation}
The true location of the peaks will depend on the energy levels of the
bound states in the full electroweak potential. For Winos and
Higgsinos with nominal mass splitting, the Coulomb resonances are
squeezed in a narrow range of velocity just below the charged
particles threshold, where the Sommerfeld factors display a
complicated pattern of peaks and dips. The peaks can be large but
are rather narrow in velocity\footnote{Here the tiny width of the resonance
  is due to decays of the Coulomb resonance to $\chi_0\chi_0$ only,
  and doesn't include decays to SM final states}. Thus we expect the physical
annihilation cross-section in the neutral channel to be enhanced at
velocities just below the charged-particle threshold.  We will come
back to this point in our discussion of the Higgsino Sommerfeld
factors below.

We can get some intuition for the Yukawa resonances by diagonalizing
the potential in the pure $SU(2)_L$ limit ($\delta m \rightarrow 0$ and $g_Y
\rightarrow 0$) to go to the isospin basis.  This yields one attractive and one repulsive
eigenvalue in each case:
\begin{equation}
V^\textrm{diag}_\textrm{doublet}=\frac{\alpha_L}{4r}e^{-M_W r}\left(\begin{array}{cc}
1 & 0\\
0 & -3
\end{array}\right)\qquad V^\textrm{diag}_\textrm{triplet}=\frac{\alpha_L}{r}e^{-M_W r}\left(\begin{array}{cc}
1 & 0\\
0 & -2
\end{array}\right)
\end{equation}
Assuming the Sommerfeld factor to be dominated by the effect of the attractive
Yukawa component, we can use the H\"ulthen potential
approximation to obtain the Sommerfeld factor in analytical form,
as well as an estimate of the mass at which the first
zero-energy bound state arises.  These are computed as
$M^*=(3/4 \alpha_L)^{-1} k M_W=5.25$ TeV ($M^*=(2 \alpha_L)^{-1} k
M_W= 1.97$ TeV) for the doublet (triplet) states; the closeness of the
latter resonance mass to that of the thermal relic Wino that
is responsible for the large enhancement to the indirect detection
signal for the Wino \cite{Cohen:2013ama,Fan:2013faa}. Instead, for the Higgsino the resonance
mass is too far away from the allowed range of relic masses to have any significant effect.  

In Figure \ref{WRes}, right panel we plot the analytical
approximation to the Wino Sommerfeld enhancement at fixed, small
velocity $\beta= 10^{-3}$  as a function of Wino mass.  For
comparison we also
show the numerical results for the full electroweak potential for
various choices of charged-neutral splitting, $\delta m$.  We see that
varying the mass splitting shifts the zero-energy resonance to
different values of the DM mass, away from the resonance position
estimated using Equation \eqref{slowYukawa}.  The dependence of the resonant mass on
the splitting can be understood using perturbation theory \cite{Bhattacharya:2018ooj}.

\begin{figure}[htb]
\begin{center}
\includegraphics[width=0.48\textwidth]{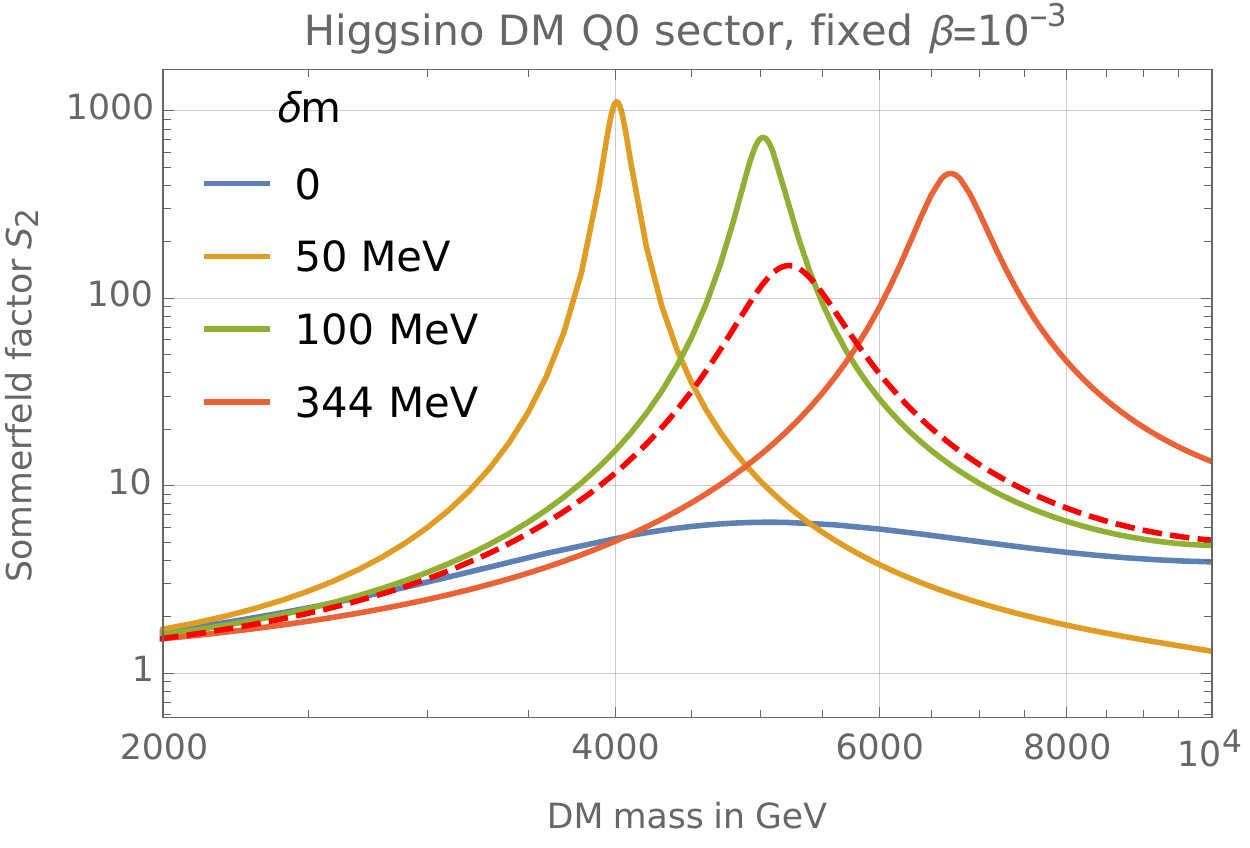} 
~
\includegraphics[width=0.48\textwidth]{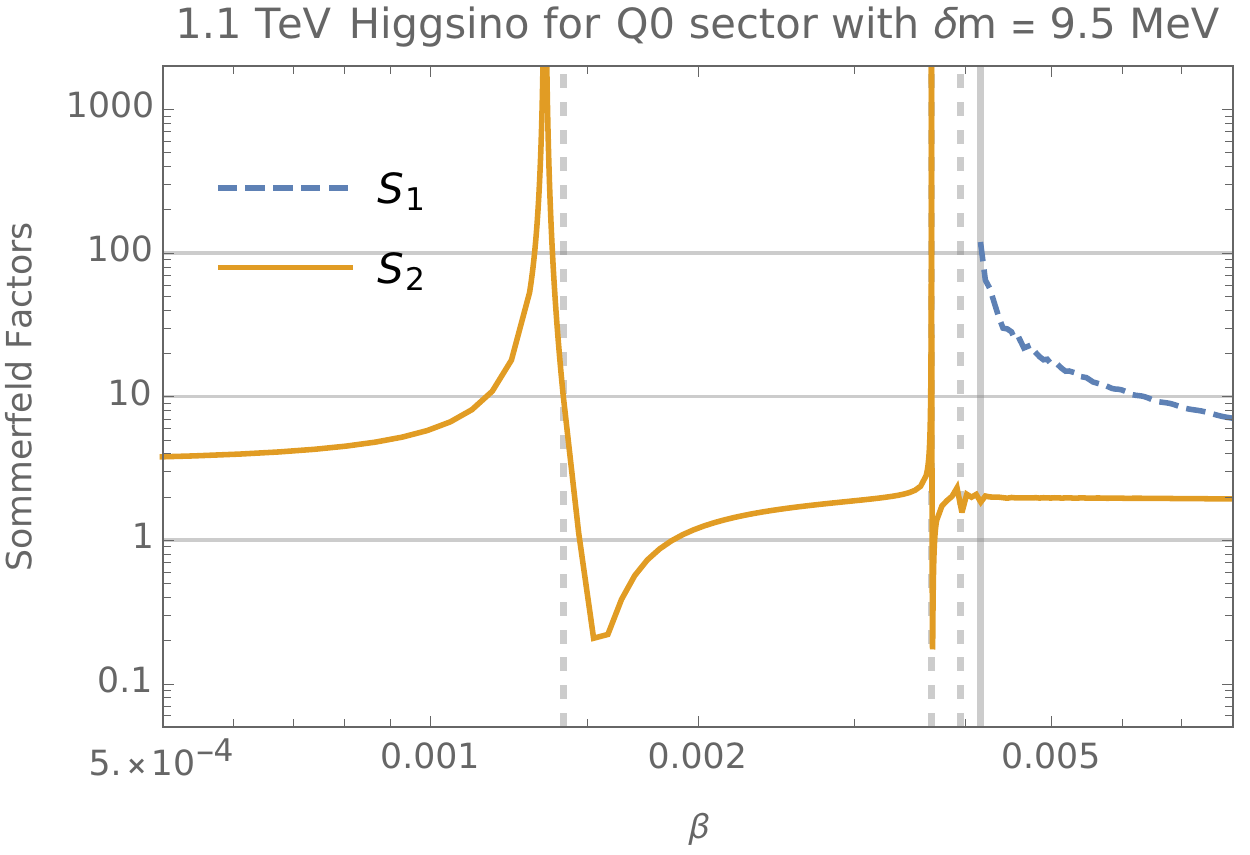} 
\end{center}
\caption{Sommerfeld factors for a pure Higgsino. Left panel: Sommerfeld factor in the neutral channel computed at fixed
  velocity ($\beta=10^{-3}$) as a function of Higgsino mass
  for varying charged-neutral splitting $\delta m$. The analytical result for
  $g_Y\to 0$ and zero splitting is shown for comparison (dashed
  line). Right
  panel: zoom around the threshold region for a 1.1~TeV thermal relic
  Higgsino with splitting $\delta m=9.5$~MeV. The solid grey line
  indicates the charged particle pair-production threshold
  $\beta_{th}$ and the grey dashed lines show the predicted positions
  of the first three Coulomb resonances.}
\label{HRes}
\end{figure}

The Sommerfeld factors for the pure Higgsino display similar features,
but the details are rather different, see Figure \ref{HRes}.  The left
panel contains the variation of the Sommerfeld factor in the
neutral-neutral channel with Higgsino mass, at $\beta= 10^{-3}$ and for varying splitting, with the analytical result
in the zero-hypercharge limit shown as a dashed red line.  In the
right panel we zoom in on the region around the charged-pair threshold
for the thermal relic Higgsino (depicted as a solid black line) with small splitting $\delta m = 9.5$
MeV, which allows us to clearly resolve the first three Coulomb
resonances.  The naive predictions for the velocities at which they
are excited (using the pure Coulomb binding
energies as in Equation \eqref{EMBS}) are marked with vertical dashed
lines, and are remarkably accurate.  For the Wino this is not the
case, the simple Coulomb approximation being off by approximately
20\%.  This is understandable in light of the larger weak Casimir factors
that the Wino is subject to, being a triplet of $SU(2)_L$.  A better
approximation for the Coulomb resonances could be obtained by
numerically solving the mixed Sommerfeld equation for the bound states.

\subsection*{Relic density}

It is well known that Sommerfeld enhancement has a significant effect on the Wino relic
density, giving rise to a global increase in the annihilation cross
section for the relic even away from the zero-energy resonance
\cite{Hisano:2006nn}.  This is not true for the relic Higgsino
\cite{Cirelli:2007xd}.  Moreover, at freeze-out velocities $x_F\sim
25$, we are always in the regime where the mass splitting is
negligible in comparison to the thermal kinetic energy $\delta m \ll
M\beta^2/2$, ensuring that the splitting, and the related Coulomb
resonances, have no effect on the thermal relic mass.  The thermal
Higgsino and Wino relic masses can thus be taken to be fixed at $1070
\pm 10$ GeV and $M=2925 \pm 15$ GeV, respectively.  We review
details of the relic density calculation including Sommerfeld
enhancement in appendix \ref{sec:Relic}.

\section{Indirect detection}
\label{sec:IndirectDetection}
Constraints from the detection of SM products of annihilating or
decaying dark matter today, or
indirect detection, come from searches for
gamma rays or antiprotons originating in dark-matter-rich regions
within and without our galaxy. In this work we focus exclusively on gamma ray
bounds, which have simpler propagation physics than the antiproton
bounds and yield comparable limits for large dark matter masses
\cite{Belanger:2012ta} and cuspy profiles \cite{Cuoco:2017iax}.  In particular we will constrain our simplified
models using measurements of the continuum photon
spectrum in the galactic centre due to HESS \cite{Abdallah:2016ygi},
and in dwarf spheroidal galaxies as given by the Fermi-LAT
collaboration \cite{Ackermann:2015zua}. 

A continuum
spectrum of photons comes mainly from the self-annihilation of DM ($\mathrm{DM} \, \mathrm{DM} \rightarrow WW + ZZ$)  and
subsequent decay and radiation of the daughter electroweak gauge bosons. 
The flux of gamma rays from DM annihilation, as seen on the earth, depend on the so-called J-factor, the integral of the DM density over the line
of sight to the source. The J-factor depends on the DM profile and its
uncertainties. From the gamma ray measurement, knowing
the spectrum of photons from the decay of $WW, ZZ$ final states and taking
into account J-factor uncertainties, one can extract bounds on the
present-day thermally-averaged annihilation cross-section $\langle \sigma \beta \rangle$
as function of the DM mass $M$. The galactic centre is
the region with the higher expected DM density and thus has a large
J-factor, however it is also home to important astrophysical gamma ray
sources.  The typical relative velocity of DM in our galaxy is
predicted to be around $\beta = 10^{-3}$. In contrast, dwarf
spheroidal satellite galaxies present less astrophysical background
but have a lower J-factor.  Furthermore the velocity of DM particles in
the dwarves halo is smaller than in our galaxy, around
$\beta=10^{-5}-10^{-4}$. Since the present day DM velocity is very low, taking into account Sommerfeld enhancement for the computation of $\langle \sigma \beta \rangle$ is crucial \cite{Fan:2013faa}. 

Today, all of the DM in the universe is neutral so the relevant
annihilation process is $\mathrm{DM}^0 \, \mathrm{DM}^0 \rightarrow WW + ZZ$.
In the absence of low-velocity resonances in the Sommerfeld factor,
the enhancement saturates at a
constant value and can be factorized out of the thermal average.  The
Sommerfeld-enhanced annihilation cross section can then be computed by
simply multiplying the tree-level cross section by the value of the
Sommerfeld factor at saturation.  Provided saturation occurs above
$\beta\sim 10^{-3}$, we will obtain equal annihilation cross
section for both the galactic centre and dwarf galaxies
\cite{Krall:2017xij}.

In the presence of Coulomb resonances however, this is no longer the
case, and we need to account for the spread in velocity of DM
particles in calculating the thermally-averaged cross section. We do
this by approximating the DM velocity as a thermal distribution,
centred at $\beta = 10^{-3}$ ($x=10^6$) for galactic signals, and
$\beta = 5 \times 10^{-5}$ ($x=4 \times 10^8$) for dwarf spheroidals.

We estimate that the annihilation cross section will receive a large enhancment in one
of two scenarios: first when the dark matter mass is close to the
resonant mass where there is a zero-energy bound state due to the
Yukawa potential (this is the case for the thermal Wino with nominal
splitting); and second, for any dark matter mass, when a
Coulomb resonance coincides with the central value for the
corresponding dark matter velocity distribution.  As argued above, in
our electroweakino examples this
can only occur when the splitting is
tuned to be of order the Coulomb binding energy.  We can use the naive
estimate in equation \eqref{EMBS} to conclude that the
indirect-detection signal due to pure Higgsinos in the galactic centre
will receive a large boost at splittings $\delta m \sim 9.0,\,2.7$
MeV, due to Coulomb resonances with $n=1,\,2$.  Similarly, solving for
$\beta\sim 5 \times 10^{-5}$ we can estimate a similar enhancement to the
indirect-detection signal in dwarf galaxies at $\delta m \sim
8.4,\,2.1$ MeV.  For larger splittings, the annihilation cross-section
to $WW$ and $ZZ$ will remain an order of magnitude below the indirect
detection bounds of HESS and Fermi-LAT \cite{Krall:2017xij}. 

\begin{figure}[tb]
\begin{center}
\includegraphics[width=0.45\textwidth]{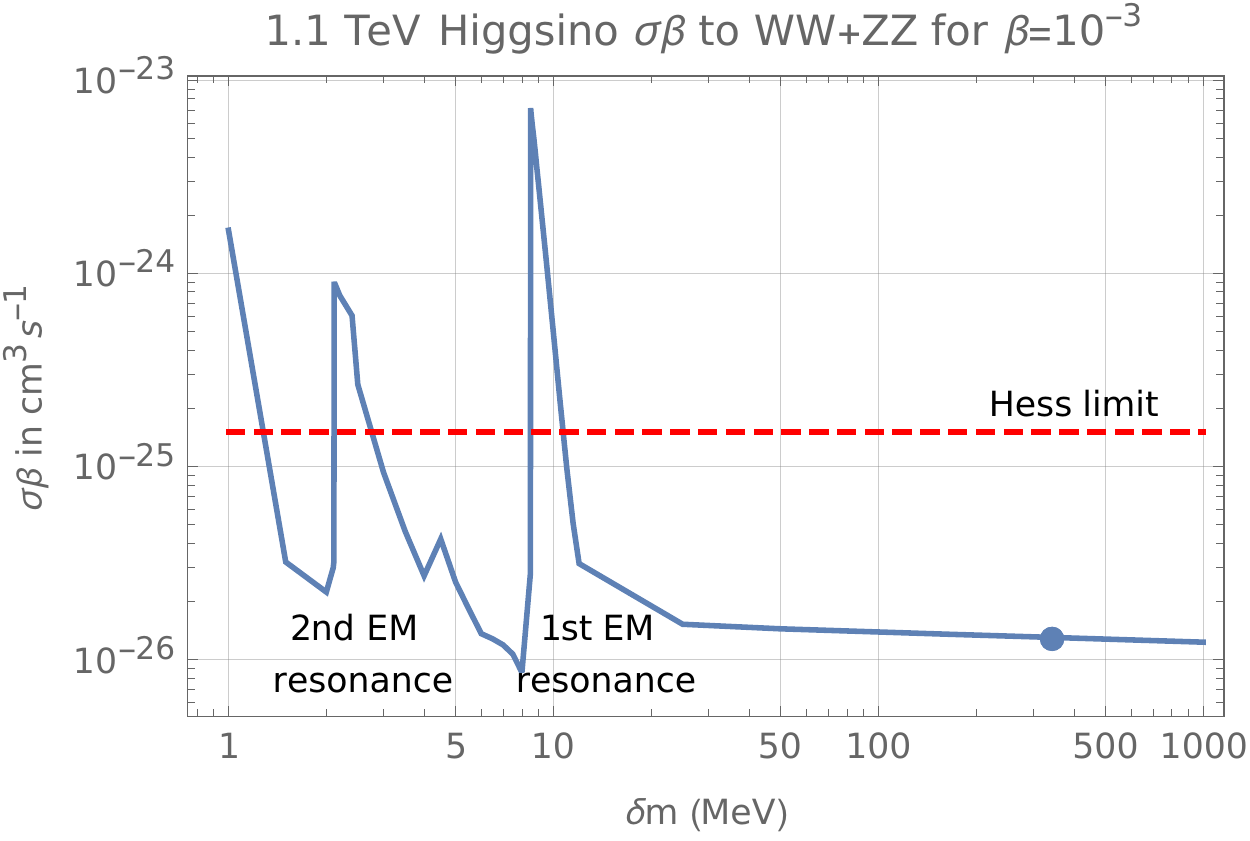} \qquad
\includegraphics[width=0.45\textwidth]{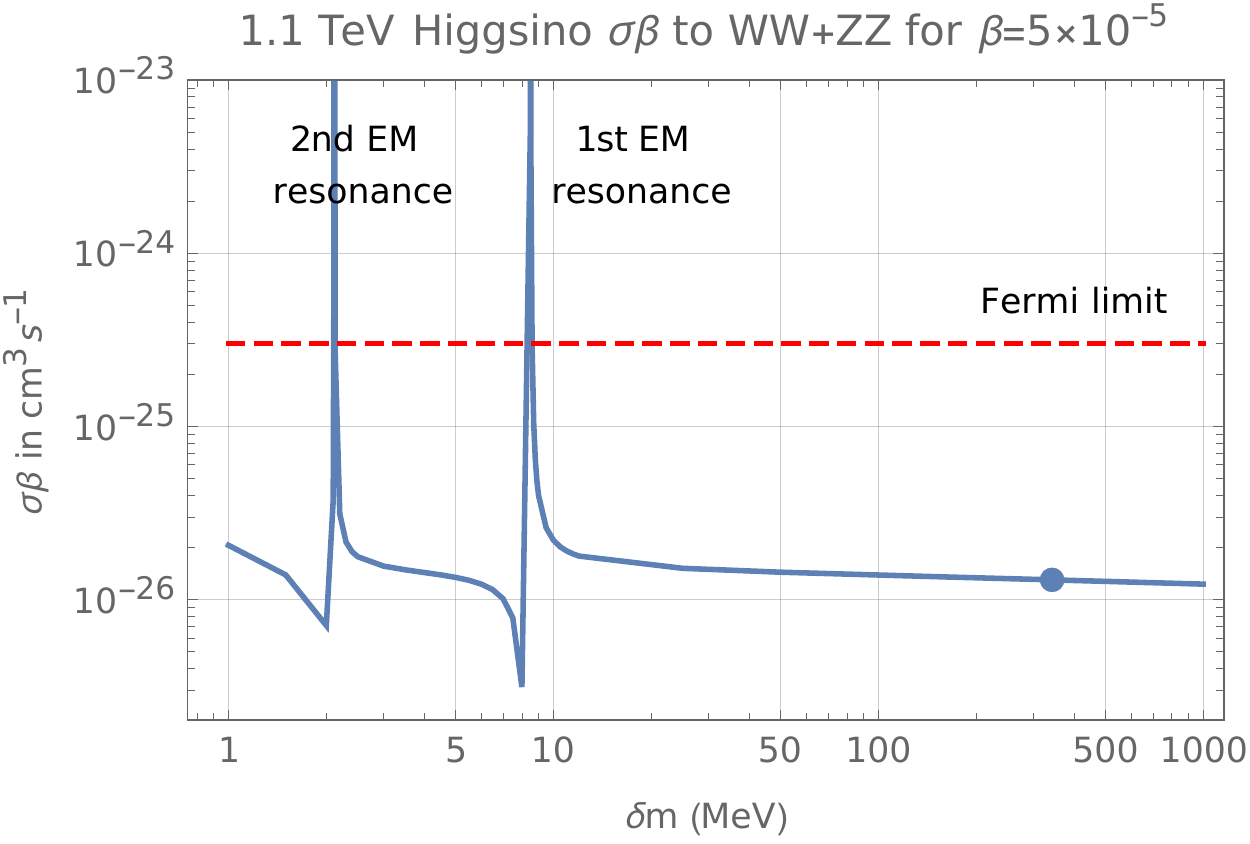}
\end{center}
\caption{Thermally-averaged annihilation cross-section to $WW+ZZ$
  for a thermal relic Higgsino as a function
  of the charged-neutral splitting for: (left panel) galactic velocities $\beta =
  10^{-3}$ and (right)  dwarf-spheroidal velocities $\beta = 5 \times
  10^{-5}$. The red dashed lines are the experimental limits due to
  HESS (left) and Fermi-LAT (right), respectively, as taken from
  \cite{Krall:2017xij}.  The nominal value for the splitting is shown
  as a large dot.}
\label{HIndirectSplit}
\end{figure}

This back-of-the-envelope estimate is borne out by the numerical
results, presented in figure \ref{HIndirectSplit} for pure Higgsino
mass close to the thermal relic value. For dark matter in
the galactic centre (left panel), the annihilation cross-section stays roughly constant as we
decrease the splitting until we reach $\delta m \sim 20$~MeV. As we
decrease the splitting still further we encounter a
very narrow peak near $\delta m = 9$~MeV that boosts the
thermally-averaged cross section by two orders of magnitude.  As we
decrease the splitting even further, the $n=1$ Coulomb
resonance crosses the $\chi^0\chi^0$ threshold and we lose the
enhancement.  The same phenomenon recurs around $\delta m=2$ Mev,
where we encounter the $n=2$ resonance.  For dwarf galaxies (right
panel) the
picture is even clearer.  The figure shows two large but very narrow
peaks slightly shifted in comparison to the ones for galactic centre
measuremnts, but compatible with the values predicted by the naive
formula.  \footnote{Although these enhancements are large, they are well within
the limit due to perturbative unitarity \cite{Blum:2016nrz}, being
naturally cut off by the total width of the resonance. Note that
in computing the Sommerfeld enhancement we only include the partial
width to dark matter final states; we are therefore neglecting the
further broadening of the resonance due to direct decays to SM final states.} From this we can conclude that for a pure Higgsino thermal
relic, charged-neutral mass splittings in the range $\delta m \in [8.5
\,, 10.5 ]$ MeV and $[2 \,, 2.5]$ MeV are excluded by measurements of the
indirect-detection cross-section by HESS.   Similarly Fermi-LAT observations exclude a very narrow
region around $\delta m = 8.5$~MeV and around $\delta m = 2.1$~MeV.
These results are relatively stable under small variations in the
Higgsino mass.

The story for the thermal Wino is rather different.  The annihilation
cross section for a pure Wino thermal relic with nominal splitting receives a significant boost
from Sommerfeld enhancement,
due to its proximity to a zero-mass resonance, bringing it well above the
bound from HESS, and comparable to that from Fermi-LAT.  Note however
that the HESS bound is computed using the Einasto profile for the halo
density, which is cuspy towards the galactic centre, and using a cored
profile would significantly relax this constraint \cite{Cirelli:2015bda}.

Decreasing the charged-neutral splitting from the nominal value should
shift the resonant mass (at which we see a zero-energy bound state) to
lower values, further away from the thermal relic mass, thus
decreasing the annihilation cross section.  As the mass splitting gets
smaller still, below $\delta m\sim 20$ MeV, as estimated from equation
\eqref{EMBS}, the first Coulomb resonance will come into play, again
resulting in an enhanced cross section.

\begin{figure}[tbh]
\begin{center}
\includegraphics[width=0.45\textwidth]{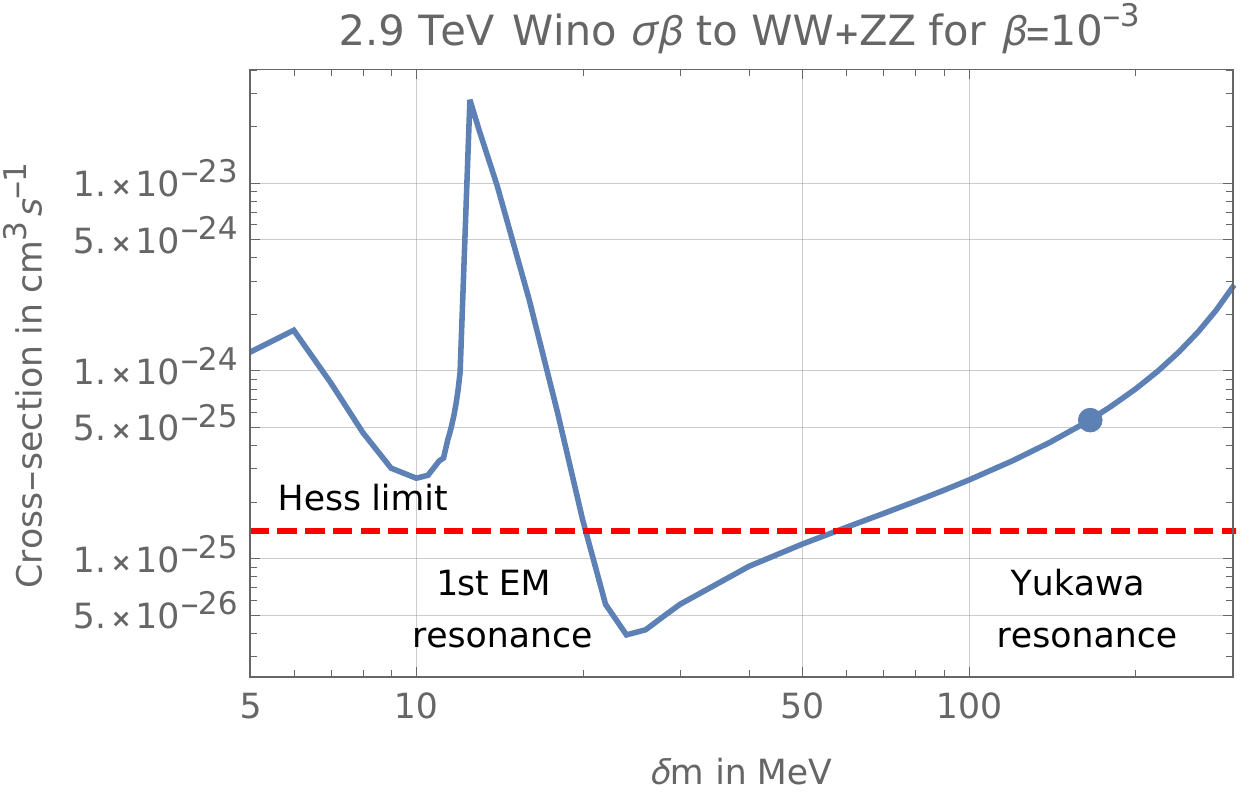} \qquad
\includegraphics[width=0.45\textwidth]{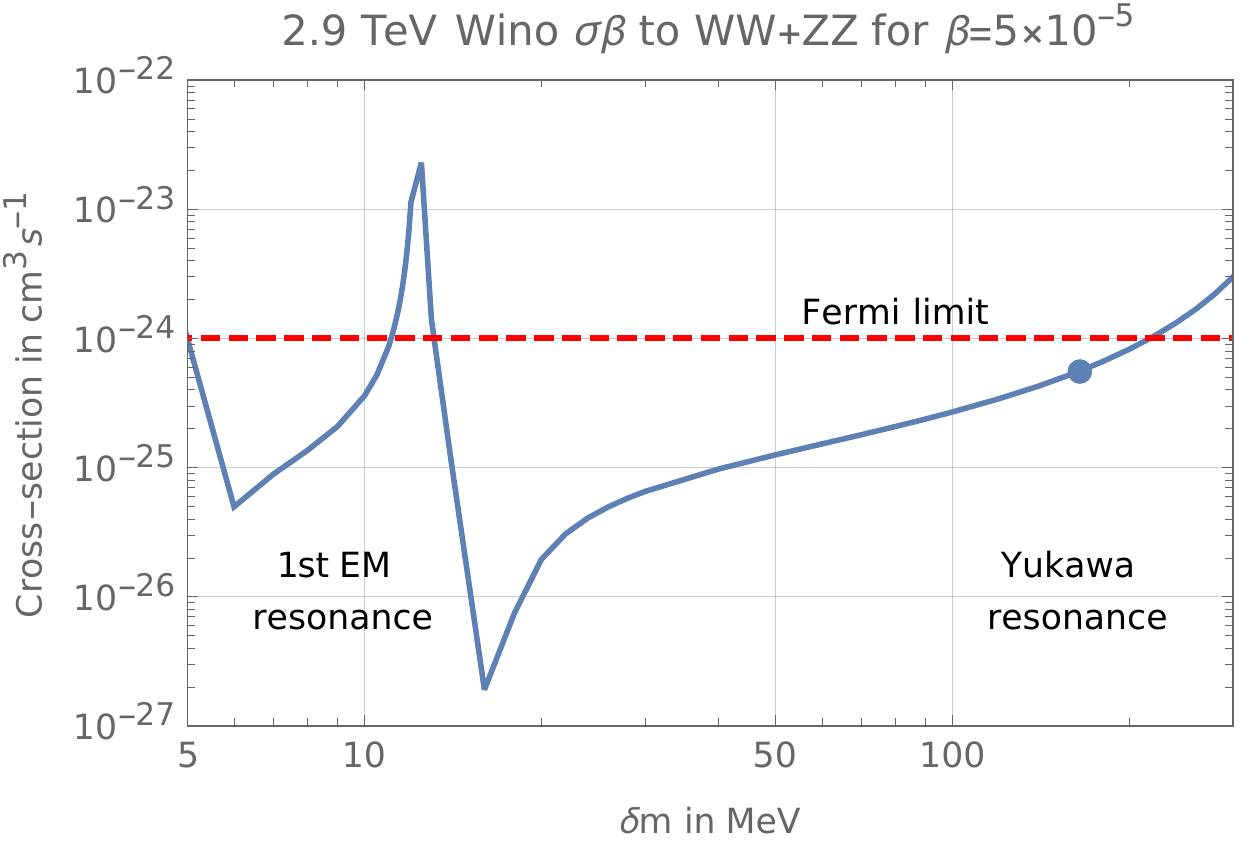}
\end{center}
\caption{Thermally-averaged annihilation cross-section to $WW+ZZ$
  for a thermal relic Wino as a function
  of the charged-neutral mass splitting for: (left panel) galactic velocities $\beta =
 10^{-3}$ and (right)  dwarf-spheroidal velocities $\beta = 5 \times
  10^{-5}$. The red dashed lines are the experimental limits due to
  HESS (left) and Fermi-LAT (right), respectively, as taken from
  \cite{Krall:2017xij}. The nominal value for the splitting is shown
  as a large dot.}
\label{IndirectSplitW}
\end{figure}

Numerical results for the thermally averaged annihilation
cross-section for pure Wino mass around the thermal relic
value are shown in figure \ref{IndirectSplitW}.  As before, dark matter today is
taken to have a thermal velocity distribution centred on $\beta=10^{-3}$ for
galactic centre measurements (left panel) and $\beta=5 \times 10^{-5}$ for
Fermi-LAT measurements (right panel). As argued above, we see a range
of intermediate splittings $\delta m \in [20,55]$ MeV
for HESS measurements and $\delta m \in [15,200]$ MeV
for Fermi-LAT, for which the annihilation cross section is smallest.
Above this value, there is a large enhancement due to a zero-energy
bound state held together by the Yukawa potential; below this value
the enhancement is due to a Coulomb resonance. In this window the
cross-section is around a factor of two below the HESS limit, well into
the uncertainty band of the J-factor \cite{Abdallah:2016ygi}. Varying the relic mass results in a small
shift of this window.

Here we tuned the splittings to these specific, small values in
order to probe regions where the effect due to Coulomb resonances is
significant.  One could instead envision a future where indirect-detection
signals are measured in galaxies and clusters at many different
scales, allowing us to `scan' over a range of dark matter velocities
in search of resonant effects.

\begin{figure}[htb]
\begin{center}
\includegraphics[width=0.45\textwidth]{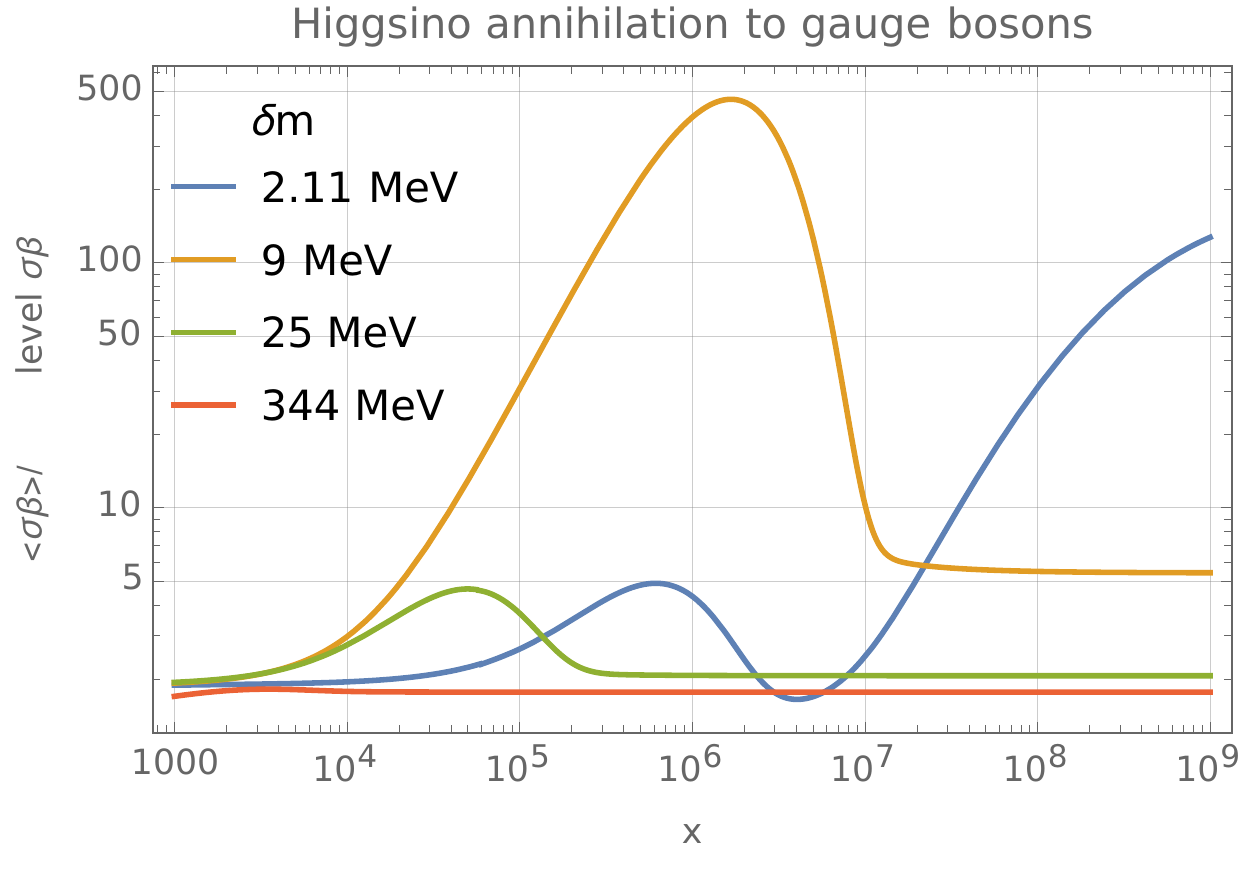}
\end{center}
\caption{Thermally-averaged annihilation cross section for 1.1~TeV
  pure Higggsino for some discrete values of the charged-neutral mass
  splitting $\delta m$.  The
  thermal average is computed using a Maxwell-Boltzman distribution
  for the dark matter velocity,
  centred at varying $x$, in order to mimic potential future measurements taken in galaxies and clusters of different
  scales.}
\label{ThermAverage}
\end{figure}
We can study this possibility by computing the annihilation
cross-section while varying the effective $x$ of the
thermal velocity distribution used in the thermal averaging procedure,
to mimic results due to dark matter in clusters of different scales.
In figure \ref{ThermAverage} we show the variation of the
sommerfeld-enhanced, thermally-averaged annihilation cross section for
a 1.1-TeV pure Higgsino with $x$ in the dark matter distribution (see
equation \eqref{equ:ThermAve}).  The same large enhancements at $\delta m =
2.1,\,9$ MeV are also visible here, at $x$ corresponding to the
average dark matter velocity in dwarf galaxies, and the galactic
centre, respectively.  As the splitting grows, we see the resonance
moving to lower values in $x$, corresponding to larger cluster sizes.
Measuring the continuum signal in galactic clusters with
$x\sim 4 \times 10^4$, for
example, would allow us to probe Higgsino mass splittings $\delta m\sim 25$
MeV.  However increasing the splitting moves the resonances to
larger velocities, resulting in smaller enhancements which can be
washed out in the thermal average. The enhancement for 25-MeV splitting is just a factor of 5 at
its maximum, whereas for nominal splitting there are many resonances
that are rather close together, and the enhancement is washed out
entirely in the thermal average.

In principle we should include thermal effects in the above
computations, the main consequence of which would be to cut off the Coulomb
enhancement in the charged channel due to the thermal mass of the
photon.  We have verified that this has a negligible effect on the
results quoted above.  Resummation of additional thermal effects that
could lead to e.g. mixing between scattering states and bound states
in the presence of a thermal plasma, have also been computed \cite{Kim:2016kxt,Binder:2018znk}, although
it is currently unclear whether these plays a significant role in
electroweakino annihilation.


\section{Two complete models}
\label{sec:FullModels}

In this section we embed the pure weakinos with variable mass
splitting in two more complete models of new physics: the MSSM, and
$N=2$ SUSY with Dirac gauginos, in order to determine the consistent
range of variation of the charged-neutral splitting $\delta m$ in
these models.

\subsection{MSSM}
The MSSM is the canonical example of a more complete
model in which the pure doublet and triplet weakinos
can be realized in different corners of its parameter space. In
addition if the weakino is the lightest supersymmetric particle (LSP),
it is automatically stable in the $R$-parity-conserving limit.  In the
fermionic sector of the MSSM, the neutral Higgsinos $\tilde H_u^0$ and
$\tilde H_d^0$ mix with the neutral electroweak gauginos $\tilde B,
\widetilde W^0$ to form four neutralinos, and the charged components
of the Higgsinos and Wino also mix, yielding two charginos. In the gauge eigenstate basis $\psi_{\tilde N} = (\tilde B, \widetilde W^0,\tilde H_u^0, \tilde H_d^0)$, the neutralino mass matrix is \cite{Martin:1997ns}:
\begin{equation}
M_{\tilde N}=
\begin{pmatrix}
M_1 & 0 & -c_\beta s_W m_Z & s_\beta s_W m_Z \\
0 & M_2 & c_\beta c_W m_Z & -s_\beta c_W m_Z \\
-c_\beta s_W m_Z & c_\beta c_W m_Z & 0 & -\mu \\
s_\beta s_W m_Z & -s_\beta c_W m_Z & -\mu & 0
\end{pmatrix}
\end{equation}
In order to avoid large CP-violating effects in the Higgs sector, we
restrict ourselves to real mass parameters $\mu, M_1, M_2$ but we
allow for one non-trivial relative sign between them.

The chargino mass matrix in the gauge-eigenstate basis $\psi_{\tilde C} = (\widetilde W^+,\tilde H_u^+,\widetilde W^-, \tilde H_d^-)$ is, in block form:
\begin{equation}
M_{\tilde C}=
\begin{pmatrix}
0 & X^T \\
X & 0
\end{pmatrix} 
\quad
\mathrm{where}
\quad
X=
\begin{pmatrix}
M_2 & \sqrt 2 s_\beta m_W \\
\sqrt 2 c_\beta m_W & \mu
\end{pmatrix} 
\end{equation}

\subsubsection*{Higgsino limit}
\label{sec:Higgsino}
In the pure Higgsino limit, $M_1, M_2 \rightarrow \infty$, the Wino
and Bino decouple giving two neutralinos and one chargino state, all with
the same tree-level mass:
\begin{equation}
m_{\tilde N_1}=m_{\tilde N_2}=m_{\tilde C_1}= \mu
\end{equation}
The two neutralino can be combined into a single Dirac fermion
Higgsino, of mass $\mu$.

Instead in the limit of large finite $M_1,\,M_2$ we can integrate out the
heavy Wino and Bino at tree-level to obtain
\begin{equation}
\mathcal{L}_{\mathrm{eff}}= \frac{g_Y^2}{4 M_1} (h_u^* \tilde H_u - h_d^* \tilde H_d)^2 + \frac{g_L^2}{4 M_2} (h_u^* \sigma^a \tilde H_u + h_d^* \sigma^a \tilde H_d)^2 + c.c.
\end{equation}
Note that this contains lepton-number-violating terms which were not
considered in the simplified model for the doublet, equation \eqref{SimplifiedHino}.
Setting the scalar Higgses to their VEVs, we get the following masses at first order in $m_W/M_{1}$, $m_W/M_{2}$:
\begin{align}
m_{\tilde N_{1,2}} =& |\mu| - \frac{\sin 2 \beta}{2} \frac{\mu}{|\mu|} \left(\frac{m_Z^2 s^2_W}{M_1} + \frac{m_W^2}{M_2} \right)   \pm \frac{1}{2} \left|\frac{m_Z^2 s^2_W}{M_1} + \frac{m_W^2}{M_2} \right| + O\left(\frac{m_W^2}{|M_1|^2} , \frac{m_W^2}{|M_2|^2} \right) \\
m_{\tilde C_1} =& |\mu| - \sin 2 \beta \frac{\mu}{|\mu|} \frac{m_W^2}{M_2}  + O\left(\frac{m_W}{|M_2|^2} \right)
\end{align} 
The tree-level mass splitting between the lightest chargino and
lightest neutralino is given by
\begin{equation} \label{MSSMsplit}
\delta m_\textrm{tree}=m_{\tilde C_1} - m_{\tilde N_{1}} =  \frac{\sin 2 \beta}{2} \frac{\mu}{|\mu|} \left(\frac{m_Z^2 s^2_W}{M_1} - \frac{m_W^2}{M_2} \right)   + \frac{1}{2} \left|\frac{m_Z^2 s^2_W}{M_1} + \frac{m_W^2}{M_2} \right| + O\left(\frac{m_W^2}{|M_1|^2} , \frac{m_W^2}{|M_2|^2} \right)
\end{equation}
There is also a splitting between the two neutral Majorana states,
generated at the same order:
\begin{equation}
\delta m_0 = M_{\tilde N_2} - M_{\tilde N_1} = m_Z^2 \left|
  \frac{s^2_W}{M_1} + \frac{c^2_W}{M_2} \right| + O\left(\frac{m_W^2}{|M_1|^2} , \frac{m_W^2}{|M_2|^2} \right)
\end{equation}
A neutral splitting  $\delta m_0>\mathcal{O}(100)$ KeV, allows the pure
the pure Higgsino to evade the strong bound from direct detection due
to $Z$-exchange.  We neglected the effect of a non-zero $\delta m_0$ in
our numerical calculations even though it is parametrically of the
same order as the charged-neutral mass splitting $\delta m$. We expect
that taking it into account will not change our main conclusion,
although it will shift the precise position of the Coulomb resonances. The effect of the neutral-neutral mass splitting on Sommerfeld enhancement is investigated in \cite{Chun:2012yt}.

We see from equation \eqref{MSSMsplit} that when $M_1$ and $M_2$ have
opposite sign, their contributions to the splitting partially cancel.
For $\mu>0$ one can have a
negative tree-level contribution to the splitting for $M_1$ negative and $M_2$ positive provided that:
\begin{equation}\label{negregion}
-M_2 \tan^2 \theta_W \frac{1+\sin 2\beta}{1-\sin 2\beta} < M_1 < -M_2 \tan^2 \theta_W \frac{1-\sin 2\beta}{1+\sin 2\beta}
\end{equation}
This region is the entire quadrant $M_1<0,\,M_2>0$ when $\tan \beta = 1$ and narrows to the line of equation $M_1 = - M_2 \tan^2 \theta_W$ as $\tan \beta$ goes to infinity. 

\begin{figure}[tbh]
\begin{center}
\begin{tabular}{c c}
   \begin{tabular}{c c}
   \includegraphics[width=0.35\textwidth]{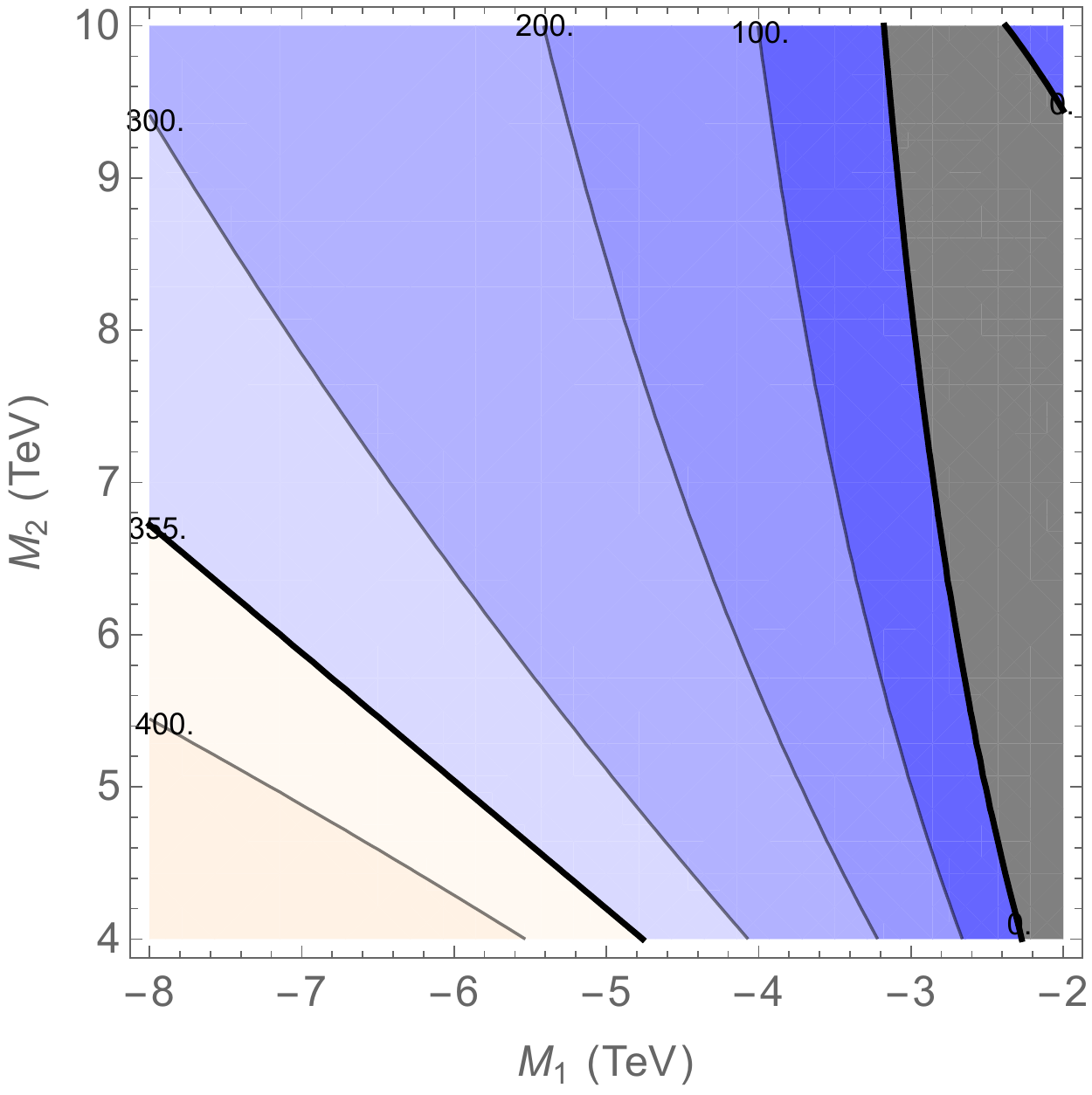}  & \includegraphics[width=0.35\textwidth]{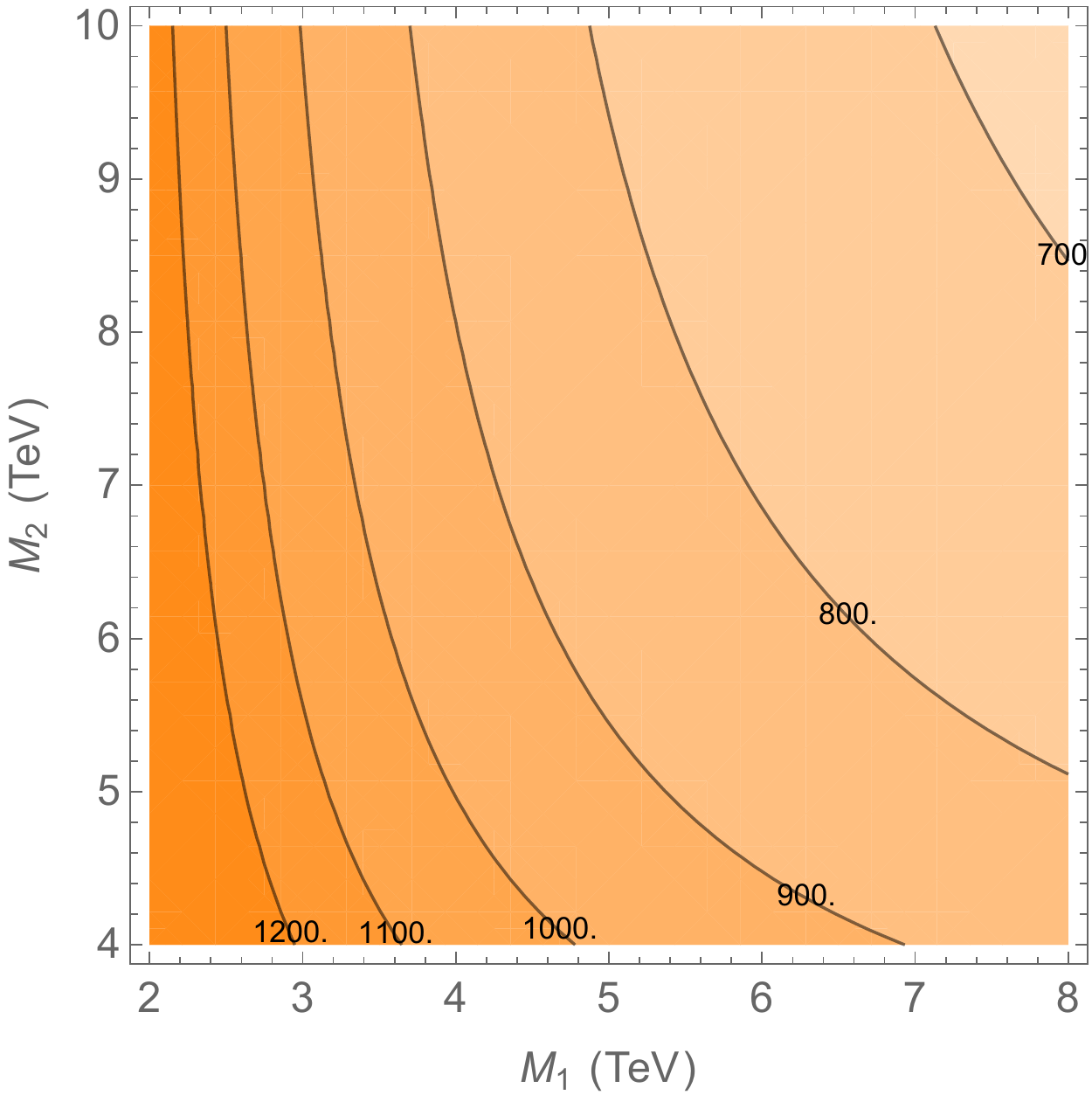} \\ 
   \includegraphics[width=0.35\textwidth]{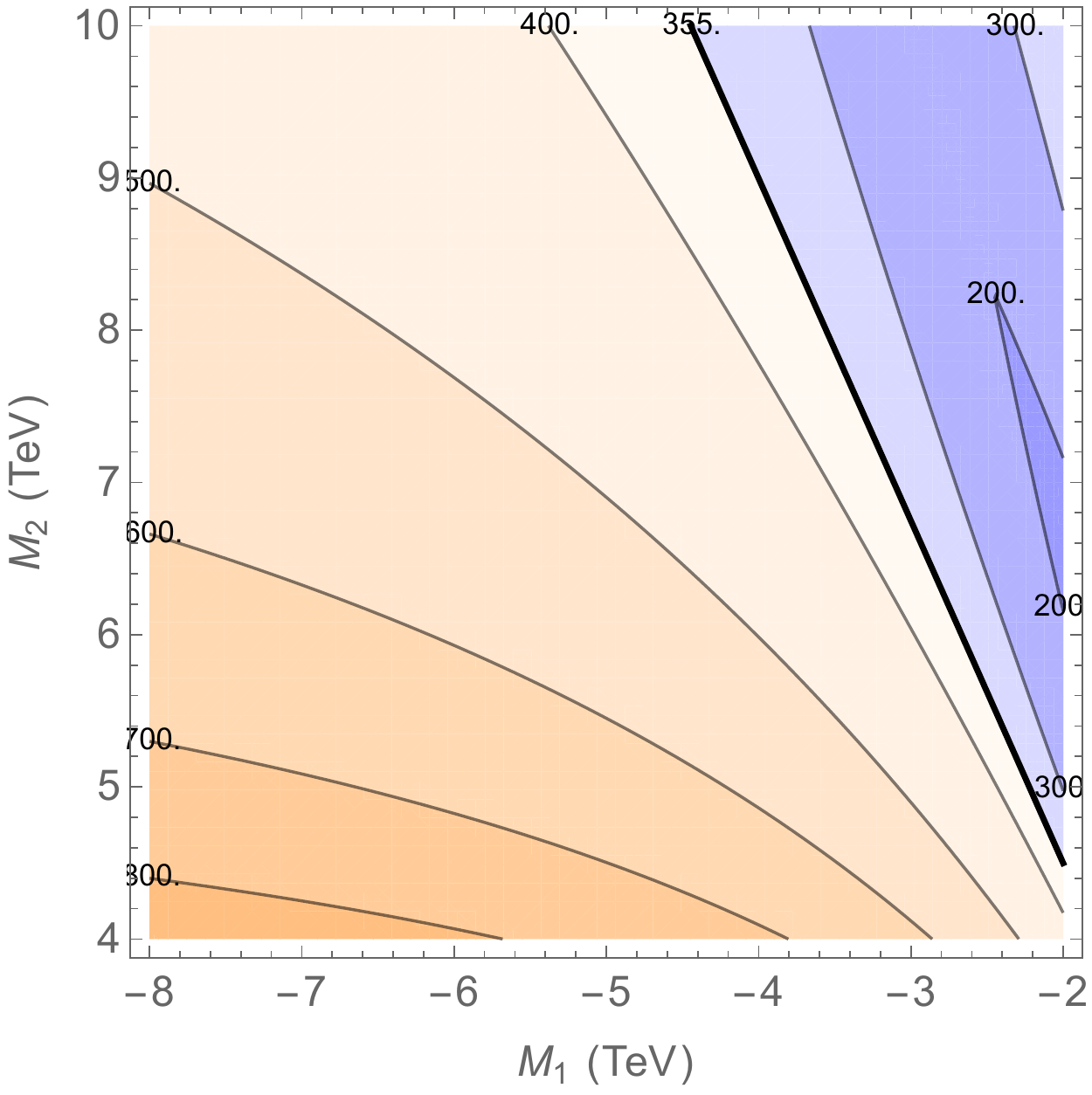} & \includegraphics[width=0.35\textwidth]{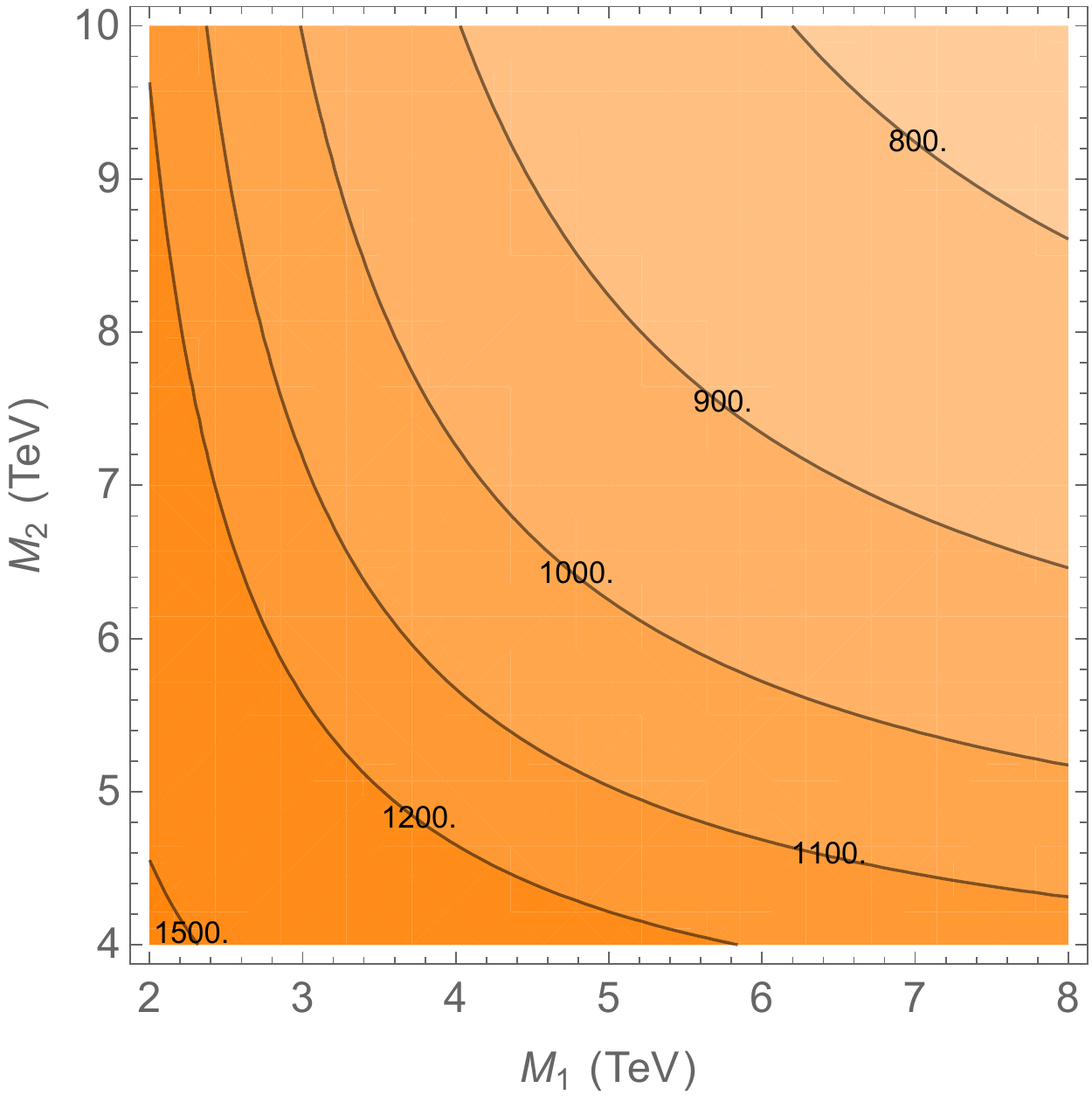} \\ 
   \includegraphics[width=0.35\textwidth]{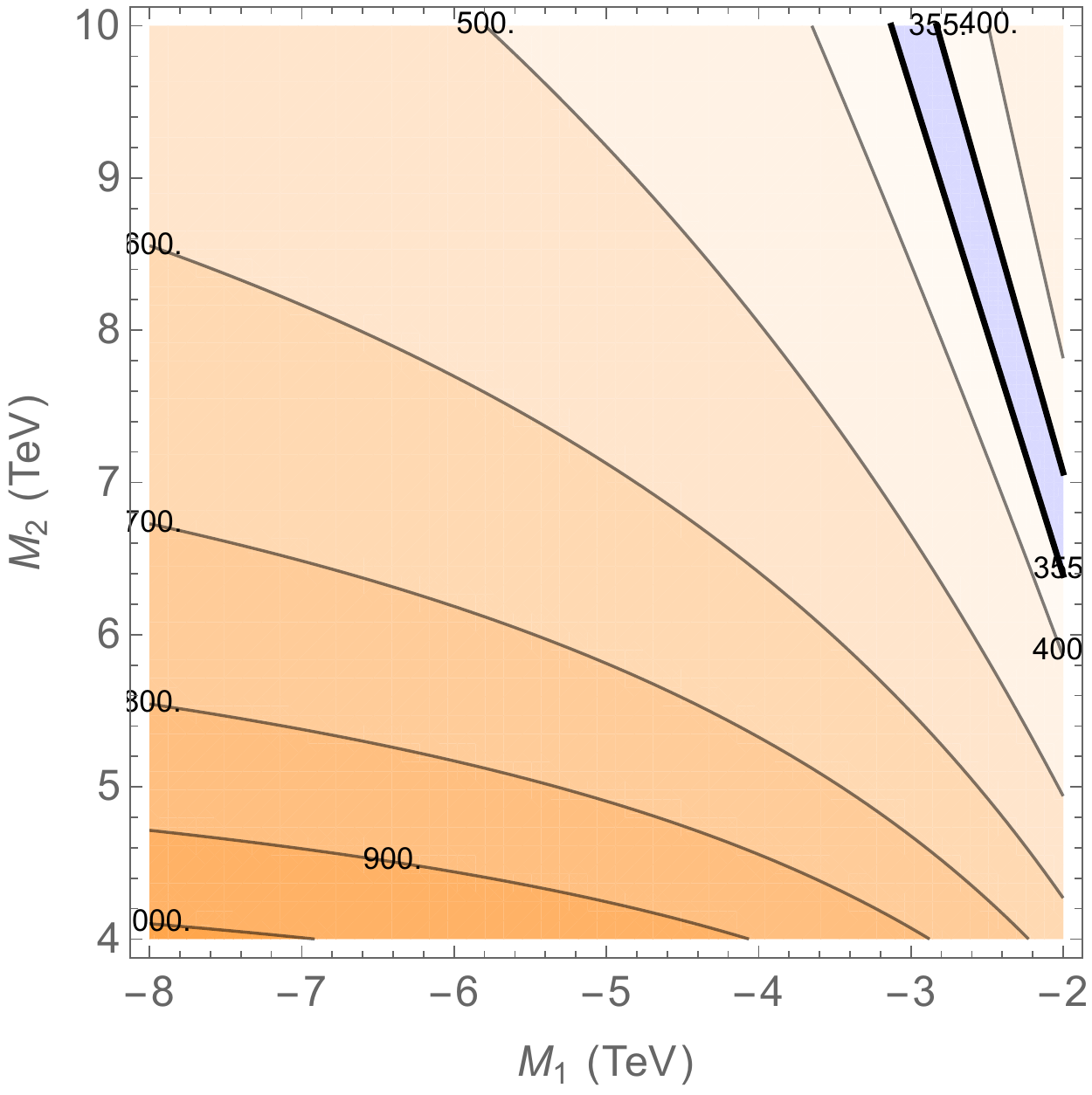} & \includegraphics[width=0.35\textwidth]{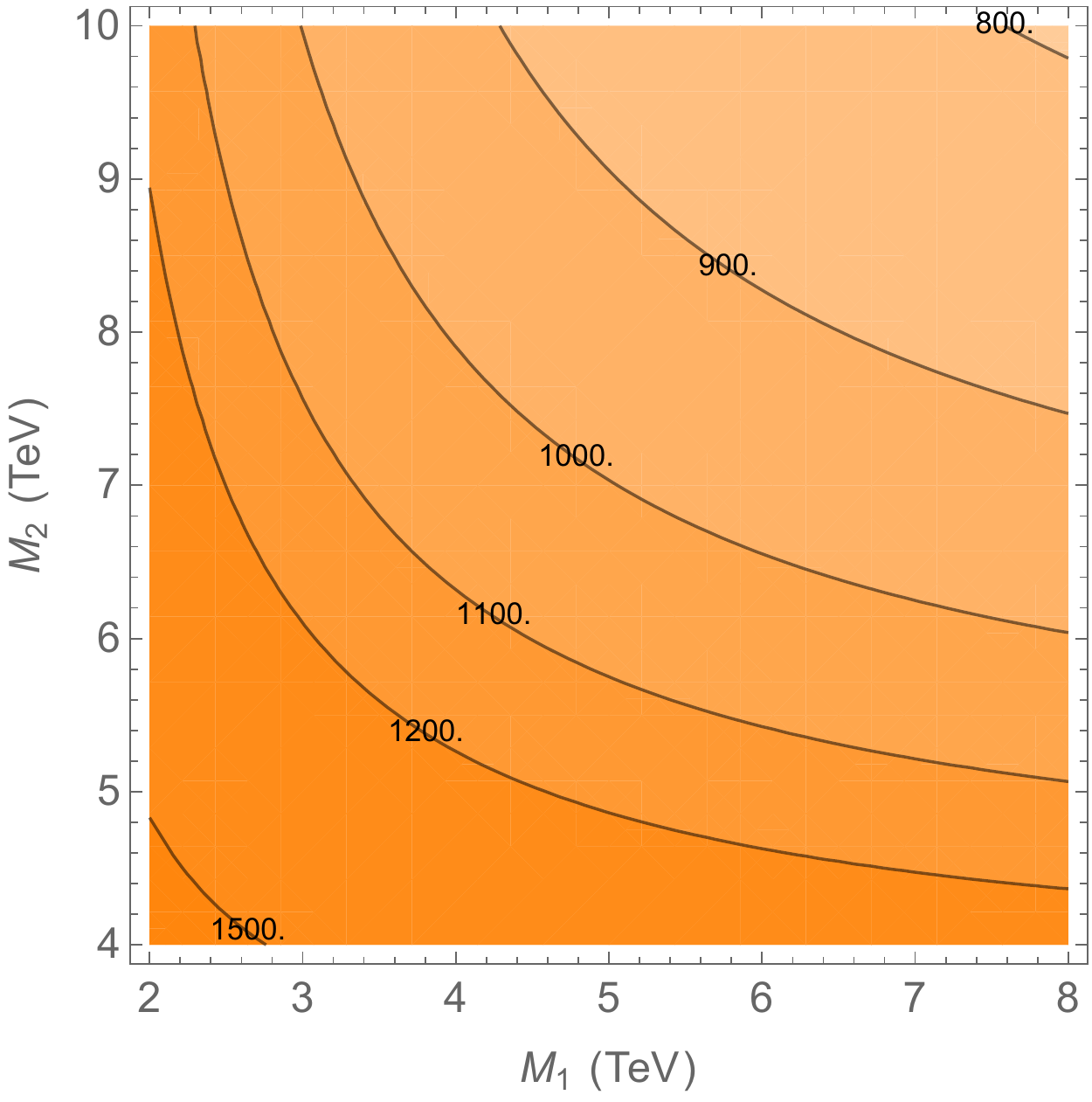} \\ 
\end{tabular}
& \begin{minipage}[c]{0.07\textwidth}
  \includegraphics[width=\textwidth]{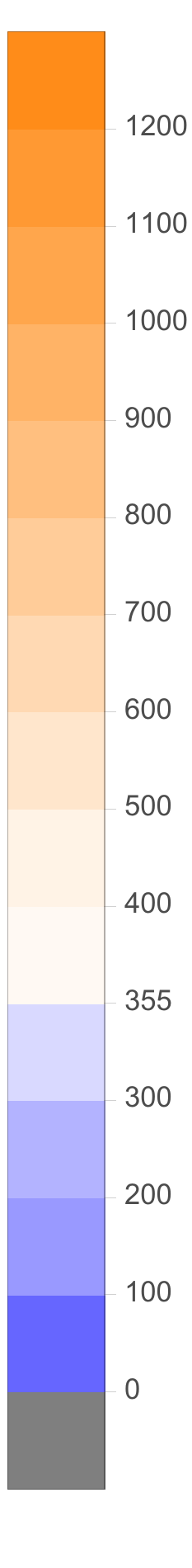} 
  \end{minipage}
\end{tabular}
\end{center}
\caption{Contours of the total charged-neutral mass splitting $\delta m$, in MeV, in the pure
  Higgsino limit of the MSSM as a function of $M_1$ and $M_2$, for
  $\tan\beta=3$ (top), $\tan\beta=10$ (middle) and $\tan\beta=80$ (bottom).}
\label{HMSSMmass}
\end{figure}

We plot the total charged-neutral mass splitting, including the
electroweak loop contribution, in figure \ref{HMSSMmass} for different
values of $\tan \beta$. As in the simplified, for fixed $M_2$ the maximum
possible range for the splitting is the interval $\delta m = \pm
\frac{m_W^2}{M_2}$, and we see that it can be much smaller than the
electroweak loop value within the MSSM parameter space, justifying our
taking $\delta m$ varying in the range $[0, \, 1.5\,\mathrm{GeV}]$.

\subsubsection*{Wino limit}
In the pure Wino limit $\mu, M_1 \rightarrow \infty$ the Bino and
Higgsino decouple from the theory and we are left with a Majorana
neutralino and a Dirac chargino that are degenerate in mass:
\begin{equation}
m_{\tilde N_1}=m_{\tilde C_1}= M_2
\end{equation}
For finite $\mu,\,M_1$ we can again integrate out the Higgsino and
Bino at tree-level to obtain
\begin{equation}
\mathcal{L}_{\mathrm{eff}}= -\frac{g_L^2}{2\mu} h_u^* \cdot h_d^* \widetilde W^a \widetilde W^a + \frac{g_L^2 g_Y^2}{2 M_1 \mu^2} \left(h_u^* \cdot (h_d^* \sigma^a) \widetilde W^a \right)^2
\end{equation}
Setting the Higgses to their VEVs, we get the following masses:
\begin{align}
m_{\tilde N_{1}} =& M_2 - \frac{m_W^2}{\mu} \sin 2\beta - \frac{m_W^4}{\mu^2 M_1} \sin^2 2\beta \tan^2 \theta_W  +\cdots\\
m_{\tilde C_1} =& M_2 - \frac{m_W^2}{\mu} \sin 2\beta +\cdots\nonumber 
\end{align} 
As in the simplified model, the mass splitting is only affected by the dimension seven operator. Explicitly:
\begin{equation}
\delta m_\textrm{tree}=m_{\tilde C_1} - m_{\tilde N_{1}} =  \frac{m_W^4}{\mu^2 M_1} \sin^2 2\beta \tan^2 \theta_W 
\end{equation}
which can be negative, for $M_1<0$.  However for any reasonable
hierarchy of masses, this tree-level contribution is
negligible compared with the electroweak loop contribution. The MSSM
in the Wino limit behaves essentially like a pure weak-triplet fermion
with charged-neutral splitting fixed at 165~MeV. As explained in
Section \ref{sec:Models} generating a splitting parametrically smaller
than the nominal value will require strong coupling at a few TeV.

\subsection{Dirac gaugino}

It is possible to obtain something closer to the weak doublet
simplified model used in this work, with a neutral splitting $\delta m_0$ that is
parametrically smaller than the charged-neutral splitting $\delta m$, in a scenario with Dirac gauginos,
coming from $N=2$ SUSY, as described in detail in
\cite{Belanger:2009wf}.  We summarize the relevant results of this
paper below.

Adding to the MSSM chiral superfields transforming in the adjoint
representation of the gauge group: a triplet $T$ for $SU(2)_L$ and a
singlet $S$ for $U(1)_Y$, allows us to write down Dirac mass terms for the gauginos:
\begin{equation}
\mathcal{L} = - m_{1D} \tilde S \tilde B - m_{2D} \tilde T^a \tilde W^a + c.c.
\end{equation}
Futhermore, we add to the MSSM superpotential all interactions that are compatible with the symmetries
\begin{equation}
W = \frac{M_S}{2} S^2 + \lambda_S\, S H_d \cdot H_u + \frac{M_T}{2} T^a T^a + \lambda_T\, T^a H_d \cdot \frac{\sigma^a}{2} H_u
\end{equation}
as well as soft masses for the adjoint superfields, forming a neutralino sector with six Majorana fermions $\tilde B, \tilde S, \widetilde W^3, \tilde T^3, \tilde H_u^0, \tilde H_d^0$.

If we take these superfields to arise from a $N=2$ supersymmetric multiplet, then the superpotential couplings take the value $\lambda_S = \sqrt{2} g_Y/2$ and $\lambda_T = \sqrt{2} g_L/2$. Keeping only the Dirac mass terms, the neutralinos form three Dirac fermions. If we now take the Higgsino limit $m_{1D}, m_{2D} \rightarrow \infty$, we are left with an Dirac electroweak doublet chargino-neutralino of mass $|\mu|$ and with mass splitting given at first order by:
\begin{equation} 
\delta m = \cos 2\beta\, m_Z^2 \left( \frac{s^2_W}{m_{1D}} - \frac{c^2_W}{m_{2D}} \right)
\end{equation}

However, if we consider that $\lambda_S$ and $\lambda_T$ take their value at some $N=2$ scale $M_{N=2}$ then the running of the couplings down to the supersymmetry breaking scale are different for $\lambda_S,\lambda_T$ than for $g_L, g_Y$ because the adjoint fields no not couple to SM matter. This generates a neutral-neutral splitting given by the following equation in the Higgsino limit  \cite{Belanger:2009wf}:
\begin{equation} 
\delta m_0 = 2 s_\beta c_\beta \mu m_Z^2 \left[\frac{c_W^2}{m_{2D}^2} \left(1-\frac{2\lambda_T^2}{g_L^2} \right) + \frac{s_W^2}{m_{1D}^2} \left(1-\frac{2\lambda_S^2}{g_Y^2} \right) \right]
\end{equation}
Neglecting all SM couplings except the top Yukawa, this is given by:
\begin{equation} 
\delta m_0 = 2 s_\beta c_\beta \mu m_Z^2 \frac{3 y_t^2}{8\pi^2}\log\left( \frac{M_{N=2}}{M_{N=1}} \right) \left[\frac{c_W^2}{m_{2D}^2} + \frac{s_W^2}{m_{1D}^2} \right]
\end{equation}
so that $\delta m_0$ is suppressed compared to $\delta m$ by a factor:
\begin{equation} 
\frac{3 y_t^2}{8\pi^2}\log\left( \frac{M_{N=2}}{M_{N=1}} \right) \left(\frac{\mu}{m_{1D,2D}}\right) \sim 10^{-2} \log\left( \frac{M_{N=2}}{M_{N=1}} \right)
\end{equation}
for $\mu=1$~TeV and $m_{1D,2D} \sim 4$~TeV. Then if we want the UV contribution to the charged neutral splitting to be of the order of the one-loop SM contribution in order to have $\delta m$ in the 10~MeV region like in Section \ref{sec:Higgsino}, the neutral-neutral splitting remains of the order of a few MeV.

\section{Conclusion}
\label{sec:Conclusion}

In this paper we explored the effect on Sommerfeld enhancement of varying the splitting between the charged and neutral components of an electroweak-multiplet dark matter candidate, focusing on pure Higgsino and Wino dark matter for concreteness, and adding to the simplified model higher-dimension operators with variable Wilson coefficient that govern the splitting.
We find as expected that the thermal relic density is unaffected by the charged-neutral splitting provided this is much smaller than the freeze-out temperature.  

For indirect detection of dark matter the story is much more interesting, as the Sommerfeld enhancement is strongly sensitive to the presence of resonances, whose effect in a multi-state system changes with the splitting.  We found that this statement doesn't only apply to the familiar `zero-energy' Yukawa resonances responsible for enhancement of the Wino indirect detection signal, but also Coulomb resonances, or quasi-bound states of $\chi^+\chi^-$, the charged components of the dark matter multiplet.  There is an infinite tower of such resonances lying between the free neutral two-particle state (ground state) and the charged particle threshold. When
the incoming dark matter particles have just enough energy to create
these states, their annihilation cross section is resonantly enhanced as in the usual Wino case.  However unlike that due to a `zero-energy' resonance, this
enhancement only occurs for certain specific values of the incoming
velocity that match the mass difference between the ground state and any of these Coulomb resonances.   

For a pure Wino/Higgsino state with nominal splitting, the Coulomb
binding energies $E_{B,\gamma}=\alpha^2M_\chi/(4n^2)$ for the charged state are negligible compared with the EW loop-induced charged-neutral splittings $
\delta m\sim\alpha\; m_Z$, making
the tower of Coulomb resonances hard to resolve, and the corresponding resonant
velocities ($\beta\approx 0.01$/$\beta\approx 0.025$) irrelevant for any measurable
physical processes.  Nevertheless in any scenario where the Coulomb
binding energy is of the order of the splitting, these resonances can
be easily resolved, and the lowest-lying resonances be brought
close enough to the ground state to give a large boost to the indirect
detection signal for dark matter today.  For pure Higgsino dark matter we found that for multiple ranges of (small) splittings, of order the
Coulomb binding energies of the $\chi^+\chi^-$ bound state, the large boost to the indirect detection
cross section at specific velocities relevant to those of dark matter
today results in strong exclusions from continuum photon
measurements, by HESS measurements in our galactic centre, and/or by
Fermi-LAT measurements in dwarf spheroidal galaxies.  We would expect
a similar enhancement to the gamma-ray line signal.  Conversely, we found that decreasing the splitting in the pure Wino scenario detunes the Yukawa resonance away from the ground state, relaxing the indirect detection constraints on the thermal Wino.

Although interesting, these results are of limited value in
themselves: decreasing the charged-neutral splitting without making
the charged state the lightest in the spectrum requires an unnatural
tuning between the contribution of a tree-level higher-dimension
operator and that due to an electroweak loop.   In the pure Wino case
there is the additional complication that the leading tree-level
operator affecting the splitting has mass dimension seven; making this
contribution of the same order as the loop correction will require
some strongly-coupled new physics.  Furthermore electroweak multiplets
with such small charged-neutral splittings will already be strongly
constrained, perhaps even excluded, both by collider searches such as
\cite{Aad:2013yna}, as well as by capture and decay/annihilation of
the heavy state in dense objects like the sun \cite{Nussinov:2009ft}.  

Instead we highlight the mechanism brought to light in this work which is rather generic, and will apply in any scenario where the dark matter two-to-two scattering contains an inelastic channel to a final state with constituents that are acted upon by a long-range force.  In analogy with the electroweakino example detailed above, this will give rise to resonances that lie below the energy threshold for final-state production due to the formation of `Coulombic' quasi-bound states, which are excited at specific incoming velocities of the annihilating dark matter, thus enhancing the annihilation cross section.  However in the pure electroweakino case both the nominal charged-neutral splitting and the Coulomb binding energies are set by the same interaction and are rather disparate in size, requiring a tuning to make the resonance velocities relevant to existing measurements.  This does not have to be the case more generally; the two relevant quantities could be unrelated, or could be naturally of the same order, which would automatically excite the Coulomb resonances at relevant velocities.

One could also invert this logic: rather than dialling the splitting between dark matter and its long-range interacting neighbour to `tune in' the dark matter annihilation signal in the particular regions of the universe where we have already made measurements, we could more ambitiously envision the universe as a dark matter spectroscope, with the different scales at which dark matter clusters today playing the part of the dial.  Measuring the indirect-detection signals in clusters at different scales would allow us to `tune' this dial in search of a signal.  If the dark matter sector contains some complex structure this should result in the spectroscope lighting up at specific `frequencies', giving us detailed information on the dark matter energy spectrum.

This work opens up various directions of study which could be fruitful
to pursue.  Further to \cite{Bhattacharya:2018ooj} it would be interesting to have an
analytical estimate of the size and widths of, and interplay between,
the different types of resonance; both the short-range Yukawa bound
states and the long-range Coulomb ones.  This would require solving
the full mixed $S=0,Q=0$ channel analytically (perhaps perturbatively) and
would yield corrections to our naive estimates for the Coulomb/Yukawa
resonance binding energies due to the other force.   We hope this would
also tell us why the analytic estimate in the $g'\,,\delta m\to 0$
limit works so well in the full mixed case for the Wino (but only for
nominal splitting), and not at all for the Higgsino.  In addition we
have not included the contribution to the annihilation of the true
formation and decay of the Coulomb bound state here.  Unlike the
Yukawa bound states, these are not the lowest-energy states in the
system, and can be produced without requiring extra radiation.
Moreover real Coulombic bound states will have additional decay
channels to Standard Model final states, that are not taken into
account here.  Finally it would be entertaining to find an alternative
dark sector framework in which the correct resonant velocities emerge
naturally.  Answering these questions would allow us to slowly piece
together the puzzle, bringing us one step closer to understanding the
nature and composition of the dark horse that is dark matter.

\section*{Acknowledgements}
We would like to give special thanks to Riccardo Rattazzi for his insightful
guidance and frequent helpful discussions. RM thanks Matthew
McCullough, John Ellis, Andrzej Hryczuk, Sasha Monin, Paolo Panci,
Michele Redi, Tracy Slatyer and Kathryn Zurek for useful
discussions, and CERN and EPFL, as well as the Galileo Galilei
Institute for Theoretical Physics for their gracious hospitality during the completion
of this project.  KM thanks Laurent Vanderheyden for his help.  RM was partially supported by the Swiss National Science
Foundation under MHV grant 171330, and also thanks the \'Etat de
Gen\`eve for its generous financial support.

\appendix
\section{Annihilation matrices}
\label{app:Gamma}
\subsection*{Weak doublet (pure Higgsino)}
Neglecting the mass splitting and all SM particle masses with respect
to the large dark matter mass $M$,  we get the following tree-level
annihilation cross-section in the center of mass frame, in the non-relativistic limit:
\begin{equation}
\sigma (\mathrm{DM} \, \overline{\mathrm{DM}} \rightarrow \mathrm{SM} ) = \frac{\pi}{256 M^2 \beta} \left[\left(81 \alpha_L^2 +12 \alpha_L \alpha_Y + 43 \alpha_Y^2\right) - \left(90 \alpha_L^2 -12 \alpha_L \alpha_Y +\frac{158}{3} \alpha_Y^2\right)\beta^2 + O(\beta^4)\right]
\end{equation} 
where $\beta$ is the dark matter velocity in the centre-of-mass frame
of the annihilating states.

For the purposes of computing the Sommerfeld enhancement, we express
the $s$-wave ($l=0$) contribution in terms of an `annihilation matrix' $\Gamma$ (the
absorptive part of the two-to-two cross section) in independent
sectors with differing electric charge:
\begin{align}
 \Gamma_{Q=0}&=\frac{\pi \alpha_L^2}{256 M^2} \begin{pmatrix}31 + 4 t_W^2 + 43 t_W^4 & -22 - 4 t_W^2 + 43 t_W^4\\ -22 - 4 t_W^2 + 43 t_W^4 & 31 + 4 t_W^2 + 43 t_W^4 \end{pmatrix} \\
 \Gamma_{Q=1}&=\frac{\pi\alpha_L^2}{128M^2}(25+4t_W^2) 
\end{align}
and also compute the nonrelativistic potential due to gauge boson
exchange in the different sectors:
\begin{align}
 V_{Q=0} &= \begin{pmatrix} 2\delta m - \alpha_{em}/r - \alpha_L (2c_W^2-1)^2 e^{-M_Z r}/4rc_W^2  & - \alpha_L e^{-M_W r}/2r \\ - \alpha_L e^{-M_W r}/2r  & - \alpha_L e^{-M_Z r}/4rc_W^2\end{pmatrix} \\
V_{Q=1} &= \frac{\alpha_L}{r}e^{-M_Z r} \frac{2c_W^2-1}{4c_W^2} 
\end{align}
where we have used the shorthand notation $c_W = \cos \theta_W$ and $t_W = \tan \theta_W$.
Then the total annihilation cross-section is given by\footnote{The
  average over the different species of DM particles and their spin is already included in the $\Gamma$ for the relic calculation.}:
\begin{equation}
\sigma \beta = \left(\sigma_{Q=0}\right)_1 + \left(\sigma_{Q=0}\right)_2 + 2 \sigma_{Q=1}
\end{equation}
where the $\sigma_Q$ are obtained by Equation \eqref{eq:SFcross} and the factor 2 comes from the counting of initial states.

Separating out the annihilation to weak gauge bosons for the purposes
of computing the indirect-detection signal:
\begin{align}
 \Gamma_{WW}= & \frac{\pi \alpha_L^2}{8 M^2}
 \begin{pmatrix} 1 & 1 \\ 1 &1  \end{pmatrix}
 \qquad
\Gamma_{ZZ}= \frac{\pi \alpha_L^2}{16 M^2 \cos^4 \theta_W}
 \begin{pmatrix} (1-2\sin^2 \theta_W)^4 & (1-2\sin^2 \theta_W)^2 \\ (1-2\sin^2 \theta_W)^2 & 1  \end{pmatrix} \\
\Gamma_{\gamma\gamma}= & \frac{\pi \alpha_\textrm{em}^2}{ M^2}
 \begin{pmatrix} 1 & 0 \\ 0 & 0  \end{pmatrix}
 \qquad
\Gamma_{\gamma Z}= \frac{\pi \alpha_L \alpha_\textrm{em}}{2 M^2 \cos^2 \theta_W}
 \begin{pmatrix} (1-2\sin^2 \theta_W)^2 & 0 \\ 0 & 0  \end{pmatrix} \nonumber
\end{align}

\subsection*{Weak triplet (pure Wino)}
For the weak triplet, the total annihilation cross section in the
nonrelativistic limit, again neglecting SM masses and inter-state mass
splittings is:
\begin{equation}
\sigma (\mathrm{DM} \, \mathrm{DM} \rightarrow \mathrm{SM} ) = \frac{\pi \alpha_L^2}{24 M^2 \beta} \left[37  - 20 \beta^2 + O(\beta^4)\right]
\end{equation}
For the purposes of the Sommerfeld computation, the $s$-wave
contribution to DM-DM annihilation can be split up into five independent
sectors with total charge $Q={0,1,2}$ and total spin $S={0,1}$
\cite{Cirelli:2007xd}. The corresponding annihilation and potential matrices are given below.
\begin{itemize}
 \item $S=0$ , $Q=0$. (This is the only sector containing mixing between
   two different two-particle states $\mathrm{DM}^+\mathrm{DM}^-$ and $\mathrm{DM}^0\mathrm{DM}^0$):
 \begin{equation}
\label{eq:GammaTotWino}
  \Gamma^{S=0}_{Q=0} = \frac{\pi \alpha_L^2}{36M^2} \begin{pmatrix} 3 & \sqrt{2} \\ \sqrt{2} & 2 \end{pmatrix}
  \qquad
  V^{S=0}_{Q=0} = \begin{pmatrix} 2\delta m - A & -\sqrt{2} B \\ -\sqrt{2} B & 0 \end{pmatrix}
 \end{equation}
\item $S=1$ , $Q=0$ ($\mathrm{DM}^+\mathrm{DM}^-$): 
 \begin{equation}
  \Gamma^{S=1}_{Q=0} = \frac{25 \pi \alpha_L^2}{144M^2} 
  \qquad
  V^{S=1}_{Q=0} = 2\delta m - A
 \end{equation}
\item $S=0$ , $Q=1$ ($\mathrm{DM}^+\mathrm{DM}^0$ and charged conjugate): 
 \begin{equation}
  \Gamma_{Q=1}^{S=0} = \frac{\pi \alpha_L^2}{36M^2} 
  \qquad
  V_{Q=1}^{S=0} = \delta m + B
 \end{equation}
 \item $S=1$ , $Q=1$ ($\mathrm{DM}^+\mathrm{DM}^0$ and charged conjugate): 
 \begin{equation}
  \Gamma_{Q=1}^{S=1} = \frac{25 \pi \alpha_L^2}{144M^2} 
  \qquad
  V_{Q=0}^{S=1} = \delta m - B
 \end{equation}
\item $S=0$ , $Q=2$ ($\mathrm{DM}^+\mathrm{DM}^+$ and charged conjugate):
  \begin{equation}
\label{eq:GammaTotWinoQ2}
  \Gamma^{S=0}_{Q=2} = \frac{\pi \alpha_L^2}{36M^2} 
  \qquad
  V^{S=0}_{Q=2} = 2\delta m + A
 \end{equation}
\end{itemize}
where we have defined:
\begin{equation}
A = \frac{\alpha_{em}}{r} + \frac{\alpha_L c_W^2}{r} e^{-M_Z r} \qquad \mathrm{and} \qquad B = \frac{\alpha_L}{r} e^{-M_W r}
 \end{equation}
Then the total annihilation to SM particles is given by:
\begin{equation}
\sigma \beta = 2\left(\sigma^{S=0}_{Q=0}\right)_1 +  \left(\sigma^{S=0}_{Q=0}\right)_2 + 2 \sigma^{S=1}_{Q=0} + 4 \sigma_{Q=1}^{S=0} + 4\sigma_{Q=1}^{S=1} + 2 \sigma^{S=0}_{Q=2}
\end{equation}
where the $\sigma^S_Q$ are given by Equation
\eqref{eq:SFcross}. The additional multiplicative factors come from the counting of initial states.

The annihilation matrices to gauge boson final states used for
computing the indirect detection cross-section are also given
in \cite{Hisano:2004ds}:
\begin{align}
 \Gamma_{WW}= \frac{\pi \alpha_L^2}{8 M^2}
 \begin{pmatrix} 2 & \sqrt 2 \\ \sqrt 2 & 4  \end{pmatrix}
 \qquad
\Gamma_{ZZ}= \frac{\pi \alpha_L^2}{2M^2}
 \begin{pmatrix} \cos^4 \theta_W & 0 \\ 0 & 0  \end{pmatrix}\\
  \Gamma_{\gamma\gamma}= \frac{\pi \alpha_{em}^2}{2M^2}
 \begin{pmatrix} 1 & 0 \\ 0 & 0  \end{pmatrix}
 \qquad
\Gamma_{\gamma Z}= \frac{\pi \alpha_L \alpha_{em}}{2M^2}
 \begin{pmatrix}2 \cos^2 \theta_W & 0 \\ 0 & 0  \end{pmatrix} \nonumber
\end{align}
These are normalized differently to the ones in \eqref{eq:GammaTotWino}-\eqref{eq:GammaTotWinoQ2}
 because only the neutral
state is present today.

\section{Relic density calculation}
\label{sec:Relic}
In the standard thermal DM scenario, the DM particles are in thermal equilibrium with the SM particles in the early universe. As the universe expands, their number density is given by the Boltzmann equation:
\begin{equation}
\frac{dn}{dt}+3Hn=-\langle \sigma \beta \rangle_{12} (n^2 - n_{eq}^2) 
\end{equation}
where $H$ is the expansion rate of the universe and $\langle \sigma
\beta \rangle_{12}$ the thermal average of the annihilation cross-section
of DM particles to SM  states multiplied by their relative velocity. 
When the temperature of the universe decreases below the DM mass $M$,
its equilibrium density drops exponentially and the annihilation rate
$\Gamma = n \langle \sigma \beta \rangle_{12}$ becomes negligible
compared to $H$. The DM particles are too rare to annihilate and their
density is only diluting with the expansion of the universe. The
temperature at which the DM particles decouple from the thermal bath is
called the freeze-out temperature $T_F$. 

Following the procedure described in \cite{Servant:2002aq}, the thermal relic abundance of DM is given by:
\begin{equation}
\Omega_{\mathrm{DM}} h^2 = \sqrt{\frac{45}{\pi}}\frac{s_0}{\rho_c} \frac{1}{\sqrt{g_*} M_P} \left[ \int_{x_F}^{\infty} \frac{\langle \sigma \beta \rangle_{12}}{x^2} dx \right]^{-1}
\end{equation}
where $s_0=2891\,\mathrm{cm}^{-3}$ is the entropy density today,
$\rho_c=1.054\times 10^{-5} h^2 \,\mathrm{GeV}\, \mathrm{cm}^3$ is the
critical density of the universe (values taken from \cite{Agashe:2014kda}), $x_F = \frac{M}{T_F}$ and $g_*$ is the number of degrees of freedom at freeze-out. Since we expect $T_F$ to be around 50 GeV we will take $g_*=92$ in all the following calculations. The freeze-out temperature is computed recursively with the formula:
\begin{equation}
x_F=\log \left[\frac{5}{4} \sqrt{\frac{45}{8}} \frac{g}{2\pi^3} \frac{M M_P}{\sqrt{g_* x_F}} \langle \sigma \beta \rangle _{T=T_F}\right]
\end{equation}
where $g$ is the number of degrees of freedom of the DM species.

With hindsight, we know for weak multiplets the mass of the
thermal relic will be of order a few TeV so we can neglect the effect of the mass splitting and consider the DM multiplet as one species. In the annihilation cross-section calculation we also neglect the splitting and we consider all SM particles as massless. We expect $x_F \sim 25$ so we will take the non-relativistic limit in the thermal average computation using the Maxwell-Boltzmann distribution:
\begin{equation}
f(\beta_{12}, x)= 4\pi \left(\frac{x}{4\pi} \right)^\frac{3}{2} \beta_{12}^2 \exp \left(-\frac{x\beta_{12}^2}{4} \right) 
\end{equation}

The annihilation cross section is often expanded in the non-relativistic limit: $\sigma \beta_{12} = a + b \beta_{12}^2 + O(\beta_{12}^4)$. In this case, the thermal average can be computed explicitly and we get:
\begin{equation}
\label{equ:ThermAve}
\int_{x_F}^{\infty} \frac{\langle \sigma \beta_{12} \rangle}{x^2} dx = \frac{1}{x_F} \left[a + \frac{3b}{x_F} + O\left(\frac{1}{x_F^2}\right) \right]
\end{equation}

We also take into account the 1-loop SM running of the $g_L$ and $g_Y$
coupling constants and evaluate these at the scale $M$,
corresponding to the approximate energy scale of the annihilation process in the
nonrelativistic limit.
%

\subsection{Pure Higgsino}
For the relic density calculation, we expect freeze-out to happen
around $x\sim 25$ and so we expect the velocity thermal distribution
to peak around $\beta_{12} \sim 0.4$. For TeV-scale  DM mass and
splittings below a GeV, annihilation occurs comfortably above the
threshold for on-shell production of the charged pair DM$^{+}$DM$^-$,
$\beta_{12,\textrm{th}}\sim 5 \times 10^{-2}$.  In this regime we
expect the potential due to the weak interaction to dominate over the
hypercharge part, and can
use the analytical computation of the Sommerfeld factor in the limit
$g_Y,\delta m\to
0$ to estimate the relic density.

In this simple limit, the potential matrix in the $Q=0$ sector can be diagonalized:
\begin{equation} \label{H0Approx}
V_{Q=0}(r)=R 
\begin{pmatrix}  \frac{1}{4}  & 0 \\ 0  & -\frac{3}{4}\end{pmatrix}
R^T \; \frac{\alpha_L}{r}  e^{-M_W r}
\quad \mathrm{where} \quad
R= \frac{1}{\sqrt{2}}\begin{pmatrix}  1  & 1 \\ -1  & 1\end{pmatrix}
\end{equation}
yielding one attractive channel and one repulsive channel.

The attractive Yukawa channel has an effective coupling of
$\frac{3}{4}\alpha_L = 0.025$ so we expect a large Sommerfeld
enhancement close to specific masses at which there is a bound state
close to threshold (zero-energy bound state), similar to the thermal wino case. The first
resonance is predicted by equation \eqref{slowYukawa} to be at
$M=\frac{4 k M_W}{3 \alpha_L} = 5.25$~TeV, far above the Higgsino DM
relic mass obtained from the perturbative calculation.\footnote{The
  positions of the first resonance with non zero $\delta m$ has to be
  computed numerically and is shown on figure \ref{HRes} (left).}

\begin{figure}[h]
\begin{center}
\includegraphics[width=0.48\textwidth]{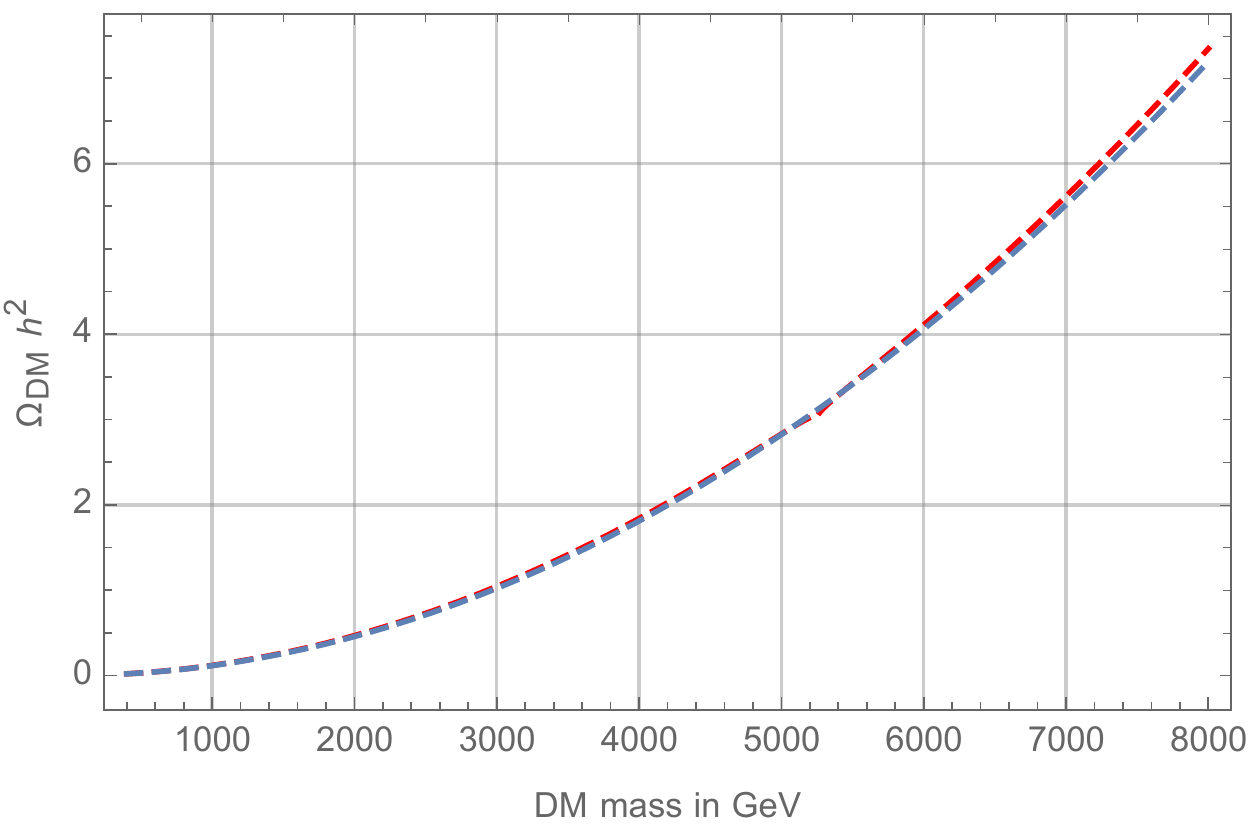} 
~
\includegraphics[width=0.48\textwidth]{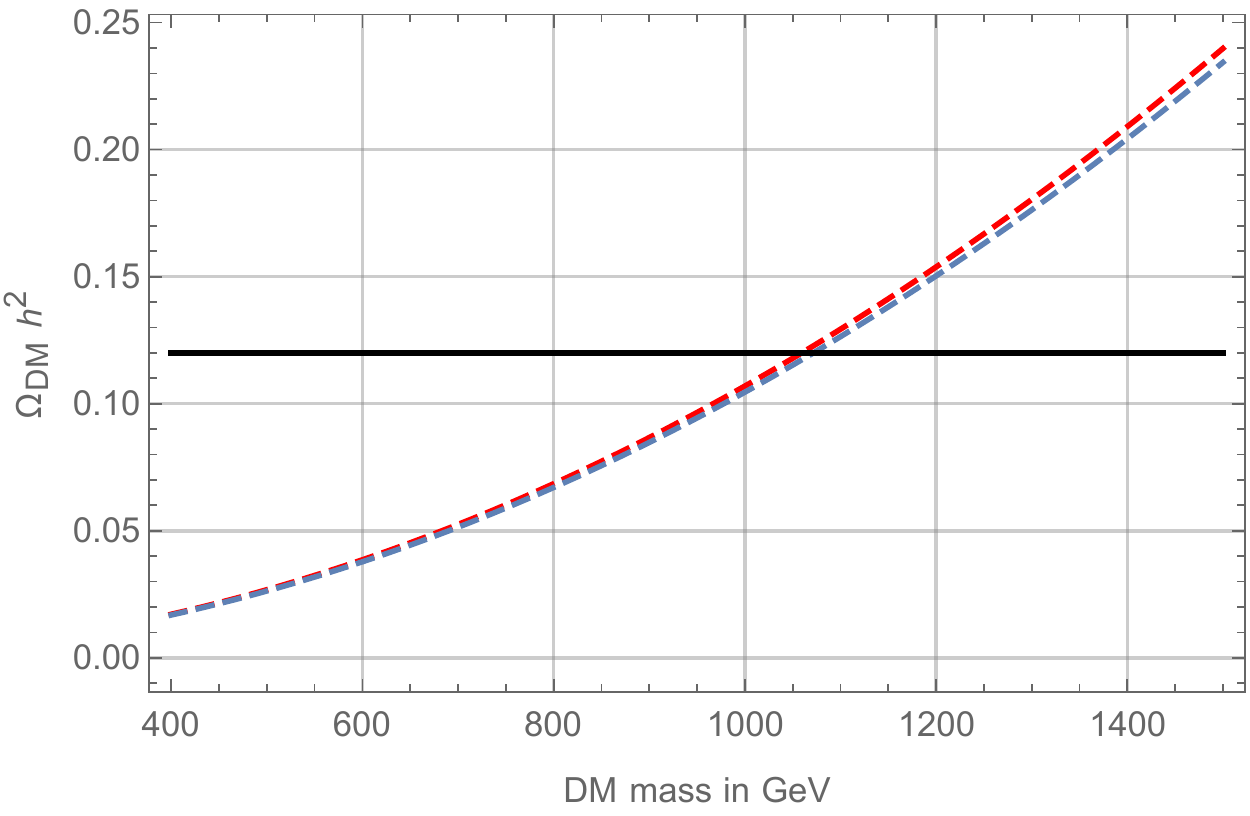} 
\end{center}
\caption{Relic density for a thermal Higgsino as a function of
  Higgsino mass. We show the standard tree-level perturbative result
  for the $l=0$ channel (blue dashed
  line), and the corresponding result including the
  non-perturbative effect computed using the analytical Sommerfeld
  factor in
  the $g_Y,\delta m\to 0$ limit (red dashed line), as described in the text.}
\label{HiggsinoAnalyticRelic}
\end{figure}

Using these approximate formulas, we can compute the Higgsino relic
density for different DM masses. The result is shown in figure
\ref{HiggsinoAnalyticRelic}; we see that Sommerfeld enhancement is
completely negligible for a wide range of masses, even close to the
resonant mass. Numerical calculations confirm that Sommerfeld
enhancement has almost no effect on nominal Higgsino DM with $\delta m
= 344$~MeV, which remains unchanged at $M=1.1$~TeV.  

For nominal splitting, the first resonance appear for a large Higgsino mass around 6.4~TeV, much above the perturbative relic mass result. But even on resonance, Sommerfeld enhancement is negligible in the relic density calculation. This can already be seen in the analytical result in figure \ref{HiggsinoAnalyticRelic}, where the resonance is predicted at 5.3~TeV but is not visible on the relic density curve.  On resonance, we see a big increase in the annihilation cross-section at very low velocity but almost no Sommerfeld enhancement at velocities relevant at freeze-out.

When decreasing the Higgsino charged-neutral splitting $\delta m$, the
resonant mass decreases, and the resonance gets sharper, as seen in figure \ref{HRes}. However, as hinted by the analytical approximation, numerical calculations show that the resonance does not affect the relic density significantly. The annihilation cross-section receives a large boost at low velocity but is almost unchanged at velocities relevant at $x_F \sim 25$ leading to very little change in the thermal average $\langle \sigma \beta_{12} \rangle$. Thus we expect the Higgsino relic mass to be insensitive to the precise value of the mass splitting. 

This will remain true provided the splitting is negligible in comparison with
the thermal kinetic energy at freeze-out, and the zero-splitting
analytical approximation holds, for $x_F \sim 25$ and $M=1.1$~TeV this
condition corresponds to $\delta m \ll 20$~GeV.  Further increasing $\delta m$
would bring the threshold \eqref{EMThresh} and the electromagnetic
resonances \eqref{EMBS} near the peak of the dark matter velocity
distribution, likely resulting in a larger enhancement of the
cross-section. This scenario will require light new physics which is
not taken into account in our simplified models, and we will not
consider it any further.


\subsection{Pure Wino}
As in the Higgsino case, at velocities relevant for the relic
calculation we expect the $SU(2)_L$ part of the potential to dominate,
and we can make an analytical estimate of the Sommerfeld factors in
the limit $g_Y,\delta m\to
0$.

In the $Q=0, S=0$ sector the potential matrix can be diagonalized:
\begin{equation}\label{W0Approx}
  V^{S=0}_{Q=0}(r) = 
  R 
\begin{pmatrix}  1  & 0 \\ 0  & -2\end{pmatrix}
R^T \; \frac{\alpha_L}{r}  e^{-M_W r}
  \quad \mathrm{where} \quad
  R = \frac{1}{\sqrt{3}}\begin{pmatrix} 1 & \sqrt{2} \\ -\sqrt{2} & 1 \end{pmatrix}
 \end{equation}
resulting again in one attractive channel and one repulsive channel. 
 
For all other channels diagonalization is not necessary, and we
simply replace compute the Sommerfeld factor replacing the Yukawa potential by the Hulth\'{e}n potential.
We estimate the resonant masses for which the Sommerfeld factors will
grow large at low velocity using equation \eqref{slowYukawa} for the
attractive channels. In the $S0Q0$ channel, the attractive Yukawa potential has an effective coupling $2 \alpha_L$ so the first resonance is predicted to be at $M=\frac{k m_W}{2 \alpha_L}=1.97$~TeV. In the $Q1$ channel, equation \eqref{slowYukawa} gives: $M=3.94$~TeV  and in the  $S1Q0$ channel $M=5.81$~TeV. The first resonance in the $S0Q0$ sector is close to the value of the Wino DM mass predicted by the perturbative calculation and is the most important for Wino DM phenomenology. The effects of a non-zero mass splitting are shown in figure \ref{WRes}; in the nominal case $\delta m =165$~MeV the resonance lies at $M=2.37$~TeV.

\begin{figure}[h]
\begin{center}
\includegraphics[width=0.48\textwidth]{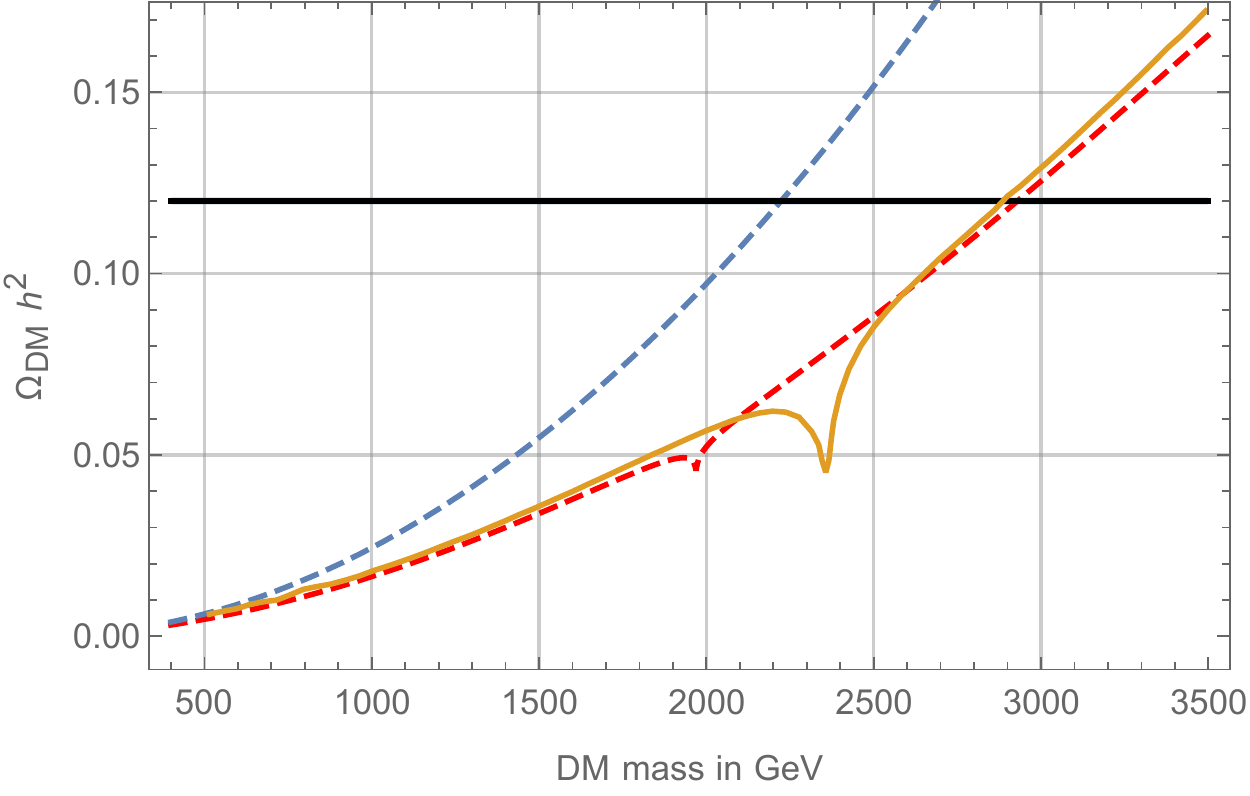} 
\end{center}
\caption{Relic density for thermal Wino DM as a function of Wino mass. We show the standard tree-level perturbative result
  for the $l=0$ channel (blue dashed
  line), and the corresponding result including the
  non-perturbative effect computed using the analytical Sommerfeld
  factor in
  the $g_Y,\delta m\to 0$ limit (red dashed line), as described in the
  text.  The solid orange line shows the numerical result extracted
  from \cite{Hisano:2006nn}, corrected to include the one-loop running
  of the gauge couplings.}
\label{WinoAnalyticRelic}
\end{figure}

We can then compute the Wino relic density as function of the DM mass. The result is shown on \ref{WinoAnalyticRelic}, we see that, unlike in the Higgsino case, Sommerfeld enhancement has an significant effect on the Wino relic density, of order 40~\%. The difference with the Higgsino scenario comes from two features of the Wino model. First in the diagonalization of the potential matrices, we see that the attractive channels have a larger effective coupling in the Wino case than in the Higgsino case resulting to overall larger Sommerfeld factors. Second, the off-diagonal terms in the annihilation matrices are negative for the Higgsino and positive for the Wino; in other words interferences between the charged and neutral channel are constructive for the Wino and destructive for the Higgsino. Moreover we see that the resonance, predicted around 2~TeV in the $\delta m =
0$ limit, has a very small effect on the relic density. The analytical approximation of Sommerfeld enhancement predicts the Wino DM mass to be $M=(2925\pm 15)$~GeV to get the correct relic abundance.

The full numerical calculation for nominal splitting $\delta m =
165$~MeV is well-approximated by the analytical result (see figure \ref{WinoAnalyticRelic}). In what follows we will take the Wino relic mass to be $M=2.9\,\mathrm{TeV}$. 

The main discrepancy between numerical and analytical results happen around the resonance in the $S0Q0$ sector. First, the position of the peak in the Sommerfeld enhancement is displaced from 1.97~TeV in the $\delta m = 0$ limit to 2.37~TeV in the nominal splitting case. Furthermore, the dip caused by the resonance is more important in the full electroweak case than in the analytical prediction. We can see on figure \ref{WinoAnalyticRelic} that the analytical formula underestimates the size of the Sommerfeld factors on resonance. At 2.9~TeV the two curves agree very well and the resonance has negligible impact on the value of the Wino relic mass.

As for the Higgsino relic, the main effect of changing the Wino mass
splitting is to shift the precise position of the resonance. This affects the Sommerfeld enhanced cross-section for small values of $\beta$ but not the high values of $\beta$ relevant for the relic density calculation. We see on figure \ref{WinoAnalyticRelic} that the analytical curve is a very good approximation for the relic density outside of the resonance dip (around $M= 2.37$~TeV) where the relic density is overestimated.  

For a thermal Wino relic we are always in the regime $\delta m \ll
T_F$ so we expect the analytic prediction (matched by the numerical
calculation for the nominal splitting $\delta m = 165$~MeV) of
$M=2.9$~TeV to be valid for all values of the mass splitting as long
as the resonance is not located right at 2.9~TeV. This happens for
$\delta m = 450$~MeV; around this value we expect the relic mass to
change by order a few hundred GeV.  Again this will require light new
states mixing with the Wino, and is outside the scope of this work.

\bibliographystyle{JHEP}
\bibliography{CoulombResonances.bib}

\providecommand{\href}[2]{#2}\begingroup\raggedright\begin{thebibliography}{10}

\bibitem{Battaglieri:2017aum}
M.~Battaglieri et~al., {\it {US Cosmic Visions: New Ideas in Dark Matter 2017:
  Community Report}},  in {\em {U.S. Cosmic Visions: New Ideas in Dark Matter
  College Park, MD, USA, March 23-25, 2017}}, 2017.
\newblock \href{http://arxiv.org/abs/1707.04591}{{\tt arXiv:1707.04591}}.

\bibitem{Baryakhtar:2017dbj}
M.~Baryakhtar, J.~Bramante, S.~W. Li, T.~Linden, and N.~Raj, {\it {Dark Kinetic
  Heating of Neutron Stars and An Infrared Window On WIMPs, SIMPs, and Pure
  Higgsinos}},  {\em Phys. Rev. Lett.} {\bf 119} (2017), no.~13 131801,
  [\href{http://arxiv.org/abs/1704.01577}{{\tt arXiv:1704.01577}}].

\bibitem{Krall:2017xij}
R.~Krall and M.~Reece, {\it {Last Electroweak WIMP Standing: Pseudo-Dirac
  Higgsino Status and Compact Stars as Future Probes}},  {\em Chin. Phys.} {\bf
  C42} (2018), no.~4 043105, [\href{http://arxiv.org/abs/1705.04843}{{\tt
  arXiv:1705.04843}}].

\bibitem{Low:2014cba}
M.~Low and L.-T. Wang, {\it {Neutralino dark matter at 14 TeV and 100 TeV}},
  {\em JHEP} {\bf 08} (2014) 161, [\href{http://arxiv.org/abs/1404.0682}{{\tt
  arXiv:1404.0682}}].

\bibitem{Mahbubani:2017gjh}
R.~Mahbubani, P.~Schwaller, and J.~Zurita, {\it {Closing the window for
  compressed Dark Sectors with disappearing charged tracks}},  {\em JHEP} {\bf
  06} (2017) 119, [\href{http://arxiv.org/abs/1703.05327}{{\tt
  arXiv:1703.05327}}]. [Erratum: JHEP10,061(2017)].

\bibitem{Sommerfeld:1931}
A.~Sommerfeld, {\it {Uber die beugung und bremsung der elektronen}},  {\em Ann.
  Phys.} {\bf 11} (1931) 257.

\bibitem{Hisano:2004ds}
J.~Hisano, S.~Matsumoto, M.~M. Nojiri, and O.~Saito, {\it {Non-perturbative
  effect on dark matter annihilation and gamma ray signature from galactic
  center}},  {\em Phys. Rev.} {\bf D71} (2005) 063528,
  [\href{http://arxiv.org/abs/hep-ph/0412403}{{\tt hep-ph/0412403}}].

\bibitem{Hisano:2006nn}
J.~Hisano, S.~Matsumoto, M.~Nagai, O.~Saito, and M.~Senami, {\it
  {Non-perturbative effect on thermal relic abundance of dark matter}},  {\em
  Phys. Lett.} {\bf B646} (2007) 34--38,
  [\href{http://arxiv.org/abs/hep-ph/0610249}{{\tt hep-ph/0610249}}].

\bibitem{ArkaniHamed:2008qn}
N.~Arkani-Hamed, D.~P. Finkbeiner, T.~R. Slatyer, and N.~Weiner, {\it {A Theory
  of Dark Matter}},  {\em Phys. Rev.} {\bf D79} (2009) 015014,
  [\href{http://arxiv.org/abs/0810.0713}{{\tt arXiv:0810.0713}}].

\bibitem{Iengo:2009ni}
R.~Iengo, {\it {Sommerfeld enhancement: General results from field theory
  diagrams}},  {\em JHEP} {\bf 05} (2009) 024,
  [\href{http://arxiv.org/abs/0902.0688}{{\tt arXiv:0902.0688}}].

\bibitem{Beneke:2014gja}
M.~Beneke, C.~Hellmann, and P.~Ruiz-Femenia, {\it {Non-relativistic pair
  annihilation of nearly mass degenerate neutralinos and charginos III.
  Computation of the Sommerfeld enhancements}},  {\em JHEP} {\bf 05} (2015)
  115, [\href{http://arxiv.org/abs/1411.6924}{{\tt arXiv:1411.6924}}].

\bibitem{Slatyer:2009vg}
T.~R. Slatyer, {\it {The Sommerfeld enhancement for dark matter with an excited
  state}},  {\em JCAP} {\bf 1002} (2010) 028,
  [\href{http://arxiv.org/abs/0910.5713}{{\tt arXiv:0910.5713}}].

\bibitem{Mitridate:2017izz}
A.~Mitridate, M.~Redi, J.~Smirnov, and A.~Strumia, {\it {Cosmological
  Implications of Dark Matter Bound States}},  {\em JCAP} {\bf 1705} (2017),
  no.~05 006, [\href{http://arxiv.org/abs/1702.01141}{{\tt arXiv:1702.01141}}].

\bibitem{Cassel:2009wt}
S.~Cassel, {\it {Sommerfeld factor for arbitrary partial wave processes}},
  {\em J. Phys.} {\bf G37} (2010) 105009,
  [\href{http://arxiv.org/abs/0903.5307}{{\tt arXiv:0903.5307}}].

\bibitem{Cirelli:2016rnw}
M.~Cirelli, P.~Panci, K.~Petraki, F.~Sala, and M.~Taoso, {\it {Dark Matter's
  secret liaisons: phenomenology of a dark U(1) sector with bound states}},
  {\em JCAP} {\bf 1705} (2017), no.~05 036,
  [\href{http://arxiv.org/abs/1612.07295}{{\tt arXiv:1612.07295}}].

\bibitem{Asadi:2016ybp}
P.~Asadi, M.~Baumgart, P.~J. Fitzpatrick, E.~Krupczak, and T.~R. Slatyer, {\it
  {Capture and Decay of Electroweak WIMPonium}},  {\em JCAP} {\bf 1702} (2017),
  no.~02 005, [\href{http://arxiv.org/abs/1610.07617}{{\tt arXiv:1610.07617}}].

\bibitem{Braaten:2017dwq}
E.~Braaten, E.~Johnson, and H.~Zhang, {\it {Zero-range effective field theory
  for resonant wino dark matter. Part III. Annihilation effects}},  {\em JHEP}
  {\bf 05} (2018) 062, [\href{http://arxiv.org/abs/1712.07142}{{\tt
  arXiv:1712.07142}}].

\bibitem{Thomas:1998wy}
S.~D. Thomas and J.~D. Wells, {\it {Phenomenology of Massive Vectorlike Doublet
  Leptons}},  {\em Phys. Rev. Lett.} {\bf 81} (1998) 34--37,
  [\href{http://arxiv.org/abs/hep-ph/9804359}{{\tt hep-ph/9804359}}].

\bibitem{Aprile:2018dbl}
{\bf XENON} Collaboration, E.~Aprile et~al., {\it {Dark Matter Search Results
  from a One Ton-Year Exposure of XENON1T}},  {\em Phys. Rev. Lett.} {\bf 121}
  (2018), no.~11 111302, [\href{http://arxiv.org/abs/1805.12562}{{\tt
  arXiv:1805.12562}}].

\bibitem{Bramante:2016rdh}
J.~Bramante, P.~J. Fox, G.~D. Kribs, and A.~Martin, {\it {Inelastic frontier:
  Discovering dark matter at high recoil energy}},  {\em Phys. Rev.} {\bf D94}
  (2016), no.~11 115026, [\href{http://arxiv.org/abs/1608.02662}{{\tt
  arXiv:1608.02662}}].

\bibitem{Cirelli:2005uq}
M.~Cirelli, N.~Fornengo, and A.~Strumia, {\it {Minimal dark matter}},  {\em
  Nucl. Phys.} {\bf B753} (2006) 178--194,
  [\href{http://arxiv.org/abs/hep-ph/0512090}{{\tt hep-ph/0512090}}].

\bibitem{Cirelli:2007xd}
M.~Cirelli, A.~Strumia, and M.~Tamburini, {\it {Cosmology and Astrophysics of
  Minimal Dark Matter}},  {\em Nucl. Phys.} {\bf B787} (2007) 152--175,
  [\href{http://arxiv.org/abs/0706.4071}{{\tt arXiv:0706.4071}}].

\bibitem{Cohen:2013ama}
T.~Cohen, M.~Lisanti, A.~Pierce, and T.~R. Slatyer, {\it {Wino Dark Matter
  Under Siege}},  {\em JCAP} {\bf 1310} (2013) 061,
  [\href{http://arxiv.org/abs/1307.4082}{{\tt arXiv:1307.4082}}].

\bibitem{Fan:2013faa}
J.~Fan and M.~Reece, {\it {In Wino Veritas? Indirect Searches Shed Light on
  Neutralino Dark Matter}},  {\em JHEP} {\bf 10} (2013) 124,
  [\href{http://arxiv.org/abs/1307.4400}{{\tt arXiv:1307.4400}}].

\bibitem{Bhattacharya:2018ooj}
A.~Bhattacharya and T.~R. Slatyer, {\it {Bound States of Pseudo-Dirac Dark
  Matter}},  \href{http://arxiv.org/abs/1812.03169}{{\tt arXiv:1812.03169}}.

\bibitem{Belanger:2012ta}
G.~Belanger, C.~Boehm, M.~Cirelli, J.~Da~Silva, and A.~Pukhov, {\it {PAMELA and
  FERMI-LAT limits on the neutralino-chargino mass degeneracy}},  {\em JCAP}
  {\bf 1211} (2012) 028, [\href{http://arxiv.org/abs/1208.5009}{{\tt
  arXiv:1208.5009}}].

\bibitem{Cuoco:2017iax}
A.~Cuoco, J.~Heisig, M.~Korsmeier, and M.~Krämer, {\it {Constraining heavy
  dark matter with cosmic-ray antiprotons}},  {\em JCAP} {\bf 1804} (2018),
  no.~04 004, [\href{http://arxiv.org/abs/1711.05274}{{\tt arXiv:1711.05274}}].

\bibitem{Abdallah:2016ygi}
{\bf H.E.S.S.} Collaboration, H.~Abdallah et~al., {\it {Search for dark matter
  annihilations towards the inner Galactic halo from 10 years of observations
  with H.E.S.S}},  {\em Phys. Rev. Lett.} {\bf 117} (2016), no.~11 111301,
  [\href{http://arxiv.org/abs/1607.08142}{{\tt arXiv:1607.08142}}].

\bibitem{Ackermann:2015zua}
{\bf Fermi-LAT} Collaboration, M.~Ackermann et~al., {\it {Searching for Dark
  Matter Annihilation from Milky Way Dwarf Spheroidal Galaxies with Six Years
  of Fermi Large Area Telescope Data}},  {\em Phys. Rev. Lett.} {\bf 115}
  (2015), no.~23 231301, [\href{http://arxiv.org/abs/1503.02641}{{\tt
  arXiv:1503.02641}}].

\bibitem{Blum:2016nrz}
K.~Blum, R.~Sato, and T.~R. Slatyer, {\it {Self-consistent Calculation of the
  Sommerfeld Enhancement}},  {\em JCAP} {\bf 1606} (2016), no.~06 021,
  [\href{http://arxiv.org/abs/1603.01383}{{\tt arXiv:1603.01383}}].

\bibitem{Cirelli:2015bda}
M.~Cirelli, T.~Hambye, P.~Panci, F.~Sala, and M.~Taoso, {\it {Gamma ray tests
  of Minimal Dark Matter}},  {\em JCAP} {\bf 1510} (2015), no.~10 026,
  [\href{http://arxiv.org/abs/1507.05519}{{\tt arXiv:1507.05519}}].

\bibitem{Kim:2016kxt}
S.~Kim and M.~Laine, {\it {On thermal corrections to near-threshold
  annihilation}},  {\em JCAP} {\bf 1701} (2017) 013,
  [\href{http://arxiv.org/abs/1609.00474}{{\tt arXiv:1609.00474}}].

\bibitem{Binder:2018znk}
T.~Binder, L.~Covi, and K.~Mukaida, {\it {Dark Matter Sommerfeld-enhanced
  annihilation and Bound-state decay at finite temperature}},  {\em Phys. Rev.}
  {\bf D98} (2018), no.~11 115023, [\href{http://arxiv.org/abs/1808.06472}{{\tt
  arXiv:1808.06472}}].

\bibitem{Martin:1997ns}
S.~P. Martin, {\it {A Supersymmetry primer}},
  \href{http://arxiv.org/abs/hep-ph/9709356}{{\tt hep-ph/9709356}}. [Adv. Ser.
  Direct. High Energy Phys.18,1(1998)].

\bibitem{Chun:2012yt}
E.~J. Chun, J.-C. Park, and S.~Scopel, {\it {Non-perturbative Effect and PAMELA
  Limit on Electro-Weak Dark Matter}},  {\em JCAP} {\bf 1212} (2012) 022,
  [\href{http://arxiv.org/abs/1210.6104}{{\tt arXiv:1210.6104}}].

\bibitem{Belanger:2009wf}
G.~Belanger, K.~Benakli, M.~Goodsell, C.~Moura, and A.~Pukhov, {\it {Dark
  Matter with Dirac and Majorana Gaugino Masses}},  {\em JCAP} {\bf 0908}
  (2009) 027, [\href{http://arxiv.org/abs/0905.1043}{{\tt arXiv:0905.1043}}].

\bibitem{Aad:2013yna}
{\bf ATLAS} Collaboration, G.~Aad et~al., {\it {Search for charginos nearly
  mass degenerate with the lightest neutralino based on a disappearing-track
  signature in pp collisions at $\sqrt(s)$=8  TeV with the ATLAS
  detector}},  {\em Phys. Rev.} {\bf D88} (2013), no.~11 112006,
  [\href{http://arxiv.org/abs/1310.3675}{{\tt arXiv:1310.3675}}].

\bibitem{Nussinov:2009ft}
S.~Nussinov, L.-T. Wang, and I.~Yavin, {\it {Capture of Inelastic Dark Matter
  in the Sun}},  {\em JCAP} {\bf 0908} (2009) 037,
  [\href{http://arxiv.org/abs/0905.1333}{{\tt arXiv:0905.1333}}].

\bibitem{Servant:2002aq}
G.~Servant and T.~M.~P. Tait, {\it {Is the lightest Kaluza-Klein particle a
  viable dark matter candidate?}},  {\em Nucl. Phys.} {\bf B650} (2003)
  391--419, [\href{http://arxiv.org/abs/hep-ph/0206071}{{\tt hep-ph/0206071}}].

\bibitem{Agashe:2014kda}
{\bf Particle Data Group} Collaboration, K.~A. Olive et~al., {\it {Review of
  Particle Physics}},  {\em Chin. Phys.} {\bf C38} (2014) 090001.

\end{thebibliography}\endgroup

\end{document}